\pdfoutput=1

\documentclass[a4paper,fleqn,usenatbib]{mnras}

\usepackage{mathptmx}

\usepackage[T1]{fontenc}
\usepackage{ae,aecompl}

\usepackage{graphicx}	
\usepackage{amsmath}	
\usepackage{amssymb}	
\usepackage{color}
\title[Inclination dependence of QPO phase lags]{Inclination dependence of QPO phase lags in black hole X-ray binaries}

\author[Van den Eijnden et al.]{
J. van den Eijnden,$^{1}$
\thanks{E-mail: a.j.vandeneijnden@uva.nl}
A. Ingram,$^{1}$
 P. Uttley,$^{1}$
 S. E. Motta,$^{2}$
 T. M. Belloni$^{3}$
 \newauthor and D. W. Gardenier$^{1}$
\\
$^{1}$Anton Pannekoek Institute for Astronomy, University of Amsterdam, Science Park 904, 1098 XH Amsterdam, The Netherlands\\
$^{2}$Department of Physics, Astrophysics, Denys Wilkinson Building, University of Oxford, Keble Road, Oxford OX1 3RH, UK\\
$^{3}$Istituto Nazionale di Astrofisica, Osservatorio Astronomico di Brera, Via E. Bianchi 46, I-23807 Merate, Italy
}

\date{Accepted XXX. Received YYY; in original form ZZZ}

\pubyear{2016}

\begin{document}
\label{firstpage}
\pagerange{\pageref{firstpage}--\pageref{lastpage}}
\maketitle

\begin{abstract}
\noindent Quasi-periodic oscillations (QPOs) with frequencies from $\sim0.05$-$30$ Hz are a common feature in the X-ray emission of accreting black hole binaries. As the QPOs originate from the innermost accretion flow, they provide the opportunity to probe the behaviour of matter in extreme gravity. In this paper, we present a systematic analysis of the inclination dependence of phase lags associated with both Type-B and Type-C QPOs in a sample of 15 Galactic black hole binaries. We find that the phase lag at the Type-C QPO frequency strongly depends on inclination, both in evolution with QPO frequency and sign. Although we find that the Type-B QPO soft lags are associated with high inclination sources, the source sample is too small to confirm this as a significant inclination dependence. These results are consistent with a geometrical origin of Type-C QPOs and a different origin for Type-B and Type-C QPOs. We discuss the possibility that the phase lags originate from a pivoting spectral power law during each QPO cycle, while the inclination dependence arises from differences in dominant relativistic effects. We also search for energy dependences in the Type-C QPO frequency. We confirm this effect in the three known sources (GRS 1915+105, H1743-322 and XTE J1550-564) and newly detect it in XTE J1859+226. Lastly, our results indicate that the unknown inclination sources XTE J1859+226 and MAXI J1543-564 are most consistent with a high inclination. 
\end{abstract}




\section{Introduction}

Black hole X-ray binary (BHXRB) systems in outburst regularly show quasi-periodic oscillations (QPOs) in their X-ray flux across a wide range of frequencies: finite-width peaks in their power spectra on top of broad band noise (BBN) \citep[see e.g.][]{vanderklis06}. BHXRBs show a large variety of QPOs, from Low-Frequency (LF) QPOs with a frequency up to $\sim 30$ Hz to High-Frequency (HF) QPOs with frequencies up to a few hundred Hz. Based on differences in, for instance, power spectral properties (such as centroid frequency, width and amplitude) and phase lag behaviour, LFQPOs can be classified into three categories: type-A, B and C \citep{wijnands99, casella05}. As type-B and type-C LFQPOs possess different properties, and have been observed simultanously in GRO J1655--40 \citep{motta12}, they are expected to originate from a different physical process. Type-A QPOs are rarely observed, and poorly understood. HFQPOs are less prevalent than LFQPOs \citep{belloni12}; in this paper, we consider only LFQPOs and will hence denote them simply as QPOs.

Despite various models, no definitive consensus about the origin of QPOs has been reached. This origin could either be \textit{geometric}, where for instance Lense-Thirring precession of the inner accretrion flow causes the observed oscillations in the X-ray flux \citep{stella98, stella99, miller05, ingram09}, or \textit{intrinsic}, where the X-ray luminosity itself is intrinsically varying. The latter could for example result from accretion rate fluctuations \citep{tagger99, cabanac10} or standing shocks in the accretion flow \citep{chakrabarti93}. A promising method to distinguish between these possible models is to search for relations between QPO properties and binary orbit inclination: \citet{schnittman06b} show that such dependencies would favor a geometric origin . 

Recently, both \citet{motta15} and \citet{heil15} have reported such an inclination dependence of QPO properties, using two completely different approaches. \citet{motta15} find that the type-C QPO shows a systematically larger absolute variability amplitude in edge-on sources, consistent with Lense-Thirring precession as its origin. The type-B QPO displays the opposite effect, having a systematically larger absolute amplitude in face-on sources, which is most consistent with an origin related to the jet. These results also support the hypothesis that these types of QPOs arise from different, but both geometric, mechanisms. Due to a small number of observations, no inclination dependence of Type-A QPOs has been detected. Secondly, \citet{heil15} show that the influence of the QPO on the BHXRB's power-colour properties is inclination dependent, implying that also the relative QPO amplitude compared to the BBN depends on inclination. 

Other evidence in support of a geometric QPO origin has also been reported recently. Using phase-resolved spectroscopy of the Type-C QPO in GRS 1915+105, \citet{ingram15} detect a modulation of the reflected iron line equivalent width, consistent with a Lense-Thirring precession model and similar earlier findings by \citet{miller05}. More recently, \citet{ingram16} also detect a modulation of the iron line centroid energy in H1743-322 using the same method. Moreover, we found that the phase lag between the hard and soft band at the Type-C QPO frequency in GRS 1915+105 systematically evolves on timescales of a few seconds \citep{vandeneijnden16}. This can be interpreted as the result of differential precession of the inner accretion flow \citep{vandeneijnden16}. Furthermore, \citet{stevens16} apply phase-resolved spectroscopy on the Type-B QPO in GX 339-4 and report quasi-periodic variation in the blackbody and power-law spectral components. This is interpreted as evidence for a jet-base origin of the Type-B QPO. Finally, \citet{homan15} show that the QPO in the neutron-star X-ray binary EXO 0748-676 can be explained by the precession of a misaligned inner accretion flow. Given this existing observational evidence for a geometric QPO origin for both Type-B and Type-C, we can expect to find more inclination dependent QPO properties. 

Since different regions in the accretion flow dominate different parts of the observed BHXRB X-ray spectrum \citep{done07}, the inclination dependence of energy-dependent QPO properties could provide new insights into the origin of type-B and type-C QPOs. One such property is the phase lag at different Fourier frequencies, in particular the QPO fundamental and (sub)harmonic, or frequencies dominated by the BBN. The different types of QPOs are known to show different phase lag behaviour at the QPO fundamental and (sub)harmonic \citep{casella05, pahari13}. Furthermore, the relation between QPO frequency and QPO phase lag differs between sources (see e.g. Figure 4 in \citealt{remillard02} and Figure 3 in \citealt{reig00}) - most noteworthy is the apparently log-linear dependence of QPO phase lags on QPO frequency in GRS 1915+105 \citep{reig00, qu10, pahari13}. These known results make phase lags an interesting QPO property to compare between sources of different inclination. 

In addition, it is also interesting to consider the inclination dependence of differences in the Type-C QPO centroid frequency in different energy bands. Such frequency differences have been reported in GRS 1915+105 \citep{qu10,yan12}, XTE J1550-564 \citep{li13} and H1743-322 \citep{li13b}. \citet{vandeneijnden16} found that the observed differences in QPO centroid frequency in GRS 1915+105 correspond to a systematic evolution of phase lags while the QPO decoheres on timescales of seconds. This is interpreted as a possible signature of differential Lense-Thirring precession in the inner accretion flow. As this interpretation proposes a geometric toy model, we also investigate the inclination dependence of the observed frequency differences in this paper.

In this paper, we present a systematic, model-independent analysis of the inclination dependence of phase lags using archival RXTE observations of fifteen BHXRBs. We determine these phase lags at various Fourier frequencies: the QPO fundamental, (sub)harmonics, and a range dominated by the broad band noise (BBN). We also measure the difference in Type-C QPO frequency between energy bands for all sources and investigate the inclination dependence of this frequency difference. We find that the phase lags show a clear dependence on inclination for the type-C QPO fundamental and BBN. Furthermore, we find significant QPO frequency differences in four sources, none of which are consistent with a low inclination. Lastly, our results indicate that XTE J1859+226 and MAXI J1543-564, both sources of undetermined inclination, are consistent with a high inclination. 

This paper is structured as follows: in Section 2, we present the source sample, the set of analysed RXTE observations, and our analysis method. In Section 3, we present our results on both phase lags and frequency differences, which are subsequently discussed in Section 4. Finally, we present our conclusions and future outlook in Section 5. 

\section{Sample and Data Analysis}
\label{sec:2}
To compare QPO and BBN phase lag properties between black hole binaries of different inclinations, we analyse archival RXTE observations of 15 Galactic BHXRBs: the fourteen sources analysed in \citet{motta15} and GRS 1915+105. For all sources except GRS 1915+105, we analyse a selection of observations based on the samples in \citet{motta15}, \citet{casella04} and \citet{st09}. We apply the same classification of sources into high (\textit{more edge-on binary orbit}), low (\textit{more face-on binary orbit}) and undetermined inclination, and of QPO observations into type-B and type-C, as \citet{motta15}. The sample includes all sources analysed by \citet{heil15} except for XTE J1118+480, GS 1354-64 and XTE J1720-318, since as these sources were not analysed by \citet{motta15}, no extensive overview of QPO observations is available. The classification of inclinations as low or high is consistent between \citet{motta15} and \citet{heil15}, except for XTE J1817-330: this source is treated as low inclination in the former, but as undetermined in the latter. 

As stated, we adopt the inclination classifications as applied in \citet{motta15}. For ten sources, a direct inclination estimate or an upper/lower limit on inclination is available. References for these estimates are listed in Table \ref{tab:sources}. Besides direct estimates, the inclination classification is mostly based on the presence of lightcurve dips, observed in all sources identified as high inclination except GRS 1915+105. Additional constraints are provided by the source's tracks in their colour-luminosity diagram, as investigated by \citet{munozdarias13}, and the observation of winds in high inclination sources. For more background on this inclination classification, we refer the reader to the extensive discussion in the main text and second appendix in \citet{motta15}.

The classification into low and high inclination sources is based on estimates of binary-orbit inclination. However, both Type-B and Type-C QPOs are thought to originate from the innermost accretion flow, which can be misaligned with the binary orbit. Hence, we adopt a broad classification, only identifying a source as either low or high inclination, instead of applying the precise estimates listed in Table \ref{tab:sources}. In Section \ref{sec:disc}, we will discuss the effect of possible inner-disk misalignments and different inclination estimates in more detail.

For GRS 1915+105, we consider the same observations as \citet{vandeneijnden16}, which are specifically selected on type-C QPOs spanning a range in frequency from $\sim 0.5$ to $\sim 8.0$ Hz. No accurate estimates of the binary orbit inclination exist for this system. However, its relativistic jet has an inclination of $70 \pm 2^{\rm o}$ \citep{mirabel94}. Assuming that any misalignment between the jet and the binary orbit axis is small, we thus classify GRS 1915+105 as a high inclination system. In Table \ref{tab:sources}, we list our complete sample, including inclination estimates and classifications, and number of type-B and type-C QPO observations. In Appendix B, we list all analysed observations with corresponding QPO frequency and measured phase lag. 

A small fraction of the observations in \citet{motta15} are not analysed in this work, for several reasons. In some cases, the QPO possesses a negligible rms amplitude at either the soft or hard energies. Alternatively, for some observations, the short exposure results in an inaccurate estimate of phase lags. In both cases, this leads to barely constrained phase lag values around the QPO fundamental and especially (sub)harmonic frequencies. In a minority of cases, we were unable to identify the QPO in the considered observation due to it possessing a low amplitude in all energy bands. In all three cases, we disregard these observations from our sample\footnote{As a result, we have not analysed any Type-C QPO observations for XTE J1817-330 and Type-B QPO observations for 4U 1630-47.}.

\begin{table}
 \begin{center}
  \caption{\small{Summary of the source sample, based on \citet{motta15}. Type-B and Type-C indicate the amount of analysed observations containing the respective QPO type. In addition to direct inclination estimates, the inclination classification is based primarily on lightcurve dips, as well as the presence of equatorial winds and the source's colour-luminosity track \citep{munozdarias13}. References for the direct inclination measure (4$^{\rm th}$ column): [1] \citet{neustroev14}, [2] \citet{orosz03}, [3] \citet{orosz04}, [4] \citet{munozdarias08,zdziarski98}, [5] \citet{millerjones11}, [6] \citet{corralsantana11}, [7] \citet{orosz11}, [8] \citet{greene01}, [9] \citet{steiner12}, [10] \citet{mirabel94}}}
  \label{tab:sources}
   \begin{tabular}{lccccc}
  \hline
  Source & C & B & $i$ & Sample & Ref. \\
  
  \hline \hline
  Swift J1753.5-01 & $31$ & -- & $\sim 40-55^{\rm o}$ & Low & [1] \\
  4U 1543-47 & $5$ & $3$ & $20.7 \pm 1.5^{\rm o}$ & Low & [2]\\
  XTE J1650-500 & $21$ & $1$ & $ > 47^{\rm o}$ & Low & [3]\\
  GX 339-4 & $46$ & $17$ & $ 40 \leq i \leq 60^{\rm o}$ & Low & [4]\\
  XTE J1752-223 & $3$ & $1$ & $\leq 49^{\rm o}$ & Low & [5]\\
  XTE J1817-330 & -- & $9$ & -- & Low & -- \\
  \hline
  XTE J1859+226 & $26$ & $14$ & $\geq 60^{\rm o}$ & -- & [6]\\
  MAXI J1543-564 & $4$ & -- & -- & -- & --\\
  \hline
  XTE J1550-564 & $63$ & $15$ & $74.7 \pm 3.8^{\rm o}$ & High & [7]\\
  4U 1630-47 & $5$ & -- & -- & High & \\
  GRO J1655-40 & $27$ & $1$ & $70.2 \pm 1^{\rm o}$ & High & [8]\\
  H1743-322 & $104$ & $36$ & $75 \pm 3^{\rm o}$& High & [9]\\
  GRS 1915+105 & $27$ & -- & $70 \pm 2^{\rm o}$ & High & [10]\\
  MAXI J1659-152 & $37$ & $6$ & -- & High & -- \\
  XTE J1748-288 & $6$ & -- & -- & High & -- \\
  \hline  
  \textit{Total} & $405$ & $103$ & & & \\
  \hline
  \end{tabular}
  \end{center}
\end{table}

Depending on the RXTE observation mode, we extract binned, event or GoodXenon lightcurves with a $1/128$ s resolution in three energy bands: a full band from $2$ to $13$ keV, and a soft and hard band from $2$ to $7$ and $7$ to $13$ keV respectively. To account for gain changes in the \textit{Proportional Counter Array} (PCA) detector, we select the absolute PCA channels most closely matching the aforementioned energy ranges for each observation individually. We calculate the power density spectrum (PDS) in the full band with a frequency resolution varying from $1/8$ Hz to $1/64$ Hz, in order to adequately sample the QPO peak while keeping the power uncertainties small. We apply an rms-squared normalisation \citep{belloni90} and fit the resulting PDS using \textsc{XSPEC} v12\footnote{\href{https://heasarc.gsfc.nasa.gov/xanadu/xspec/}{https://heasarc.gsfc.nasa.gov/xanadu/xspec/}} with a model consisting of a constant white noise and a combination of Lorentzians to account for the BBN, the QPO fundamental and possible QPO (sub)harmonic. 

To calculate phase lags, we determine the complex cross spectrum between the soft and hard band lightcurve and average the real and imaginary part over the considered frequency ranges. We then calculate the phase lag from this averaged cross spectrum, while we determine the errors through the raw coherence \citep[see e.g.][section 2]{uttley14}. The cross spectrum at the QPO fundamental and (sub)harmonic are averaged over a symmetric range around the fitted Lorentzian centroid frequency with a width of the fitted FWHM. The BBN cross spectrum is averaged over the frequency range between $0.5$ and $1.5$ Hz, provided that both the QPO and subharmonic frequencies are larger than $2$ Hz to ensure that the lags are not directly contaminated by the QPO. In this paper, following common convention, we define hard lags (i.e. hard photons lagging soft photons) as positive. 

In addition to phase lags between a broad soft and hard band, we also calculate lag-energy spectra of the Type-C QPO: the Type-C QPO lag between narrow bands and a soft reference band, as function of energy. For this purpose, we extract lightcurves at the highest possible energy resolution given the observation's data mode. We calculate the cross spectrum between each narrow energy band lightcurve and the softest energy band lightcurve (generally between $\sim 2$ and $\sim 3-4$ keV, depending on the RXTE data mode), using the same method as detailed above \citep[see][for a more extensive description of this procedure]{uttley14}. Subsequenctly, we average the real and imaginary part of the cross spectra at the QPO frequency range in each cross spectrum, calculate the QPO lag from this averaged cross spectrum, and plot the lag as a function of energy. As we will discuss in the next section, these lag-energy spectra show merely broad features. This shows that our selection of broad energy bands does not wash out any narrow lag features.

Finally, for each Type-C QPO observation, we also calculate the difference between the QPO centroid frequency in the broad hard and soft band. We calculate the normalised PDS of the two broad energy band lightcurves, and fit the QPO frequency for each lightcurve using the multi-Lorentzian fit detailed above. We define the frequency difference $\Delta \nu_0$ as the QPO frequency in the hard band minus the QPO frequency in the soft band. When plotting this frequency difference, we rescale it by the QPO frequency in the full energy band $\nu_0$.

\section{Results}

\begin{figure}
  \begin{center}
    \includegraphics[width=\columnwidth]{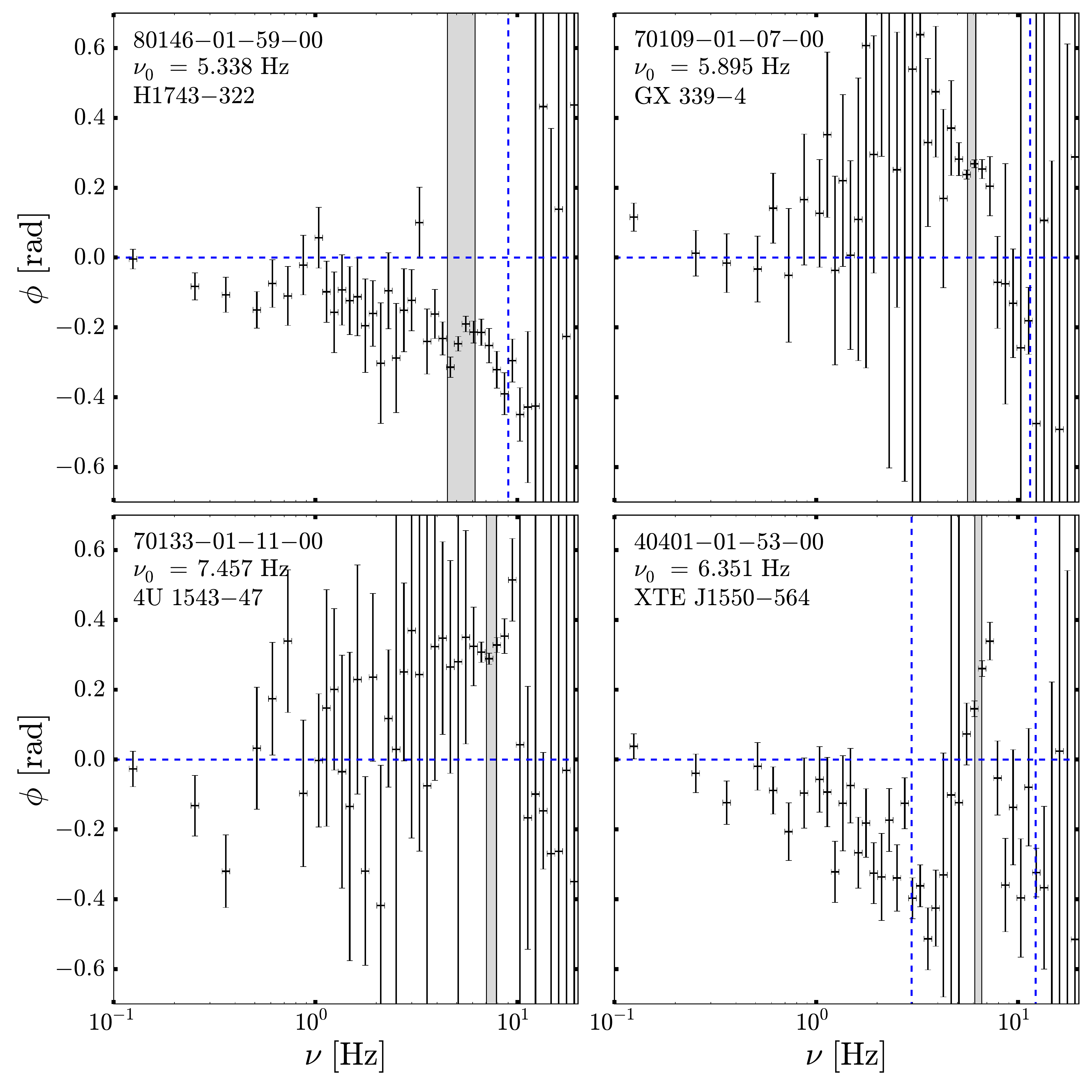}
    \caption{Examples of geometrically binned Type-B QPO lag-frequency spectra. The grey band indicates the range over which the QPO lag is averaged ($\nu_0 \pm \text{FWHM}/2$). The vertical dotted lines indicate (sub)harmonic frequencies.}
    \label{fig:LS_B}
  \end{center}
\end{figure}
\begin{figure}
  \begin{center}
        \includegraphics[width=\columnwidth]{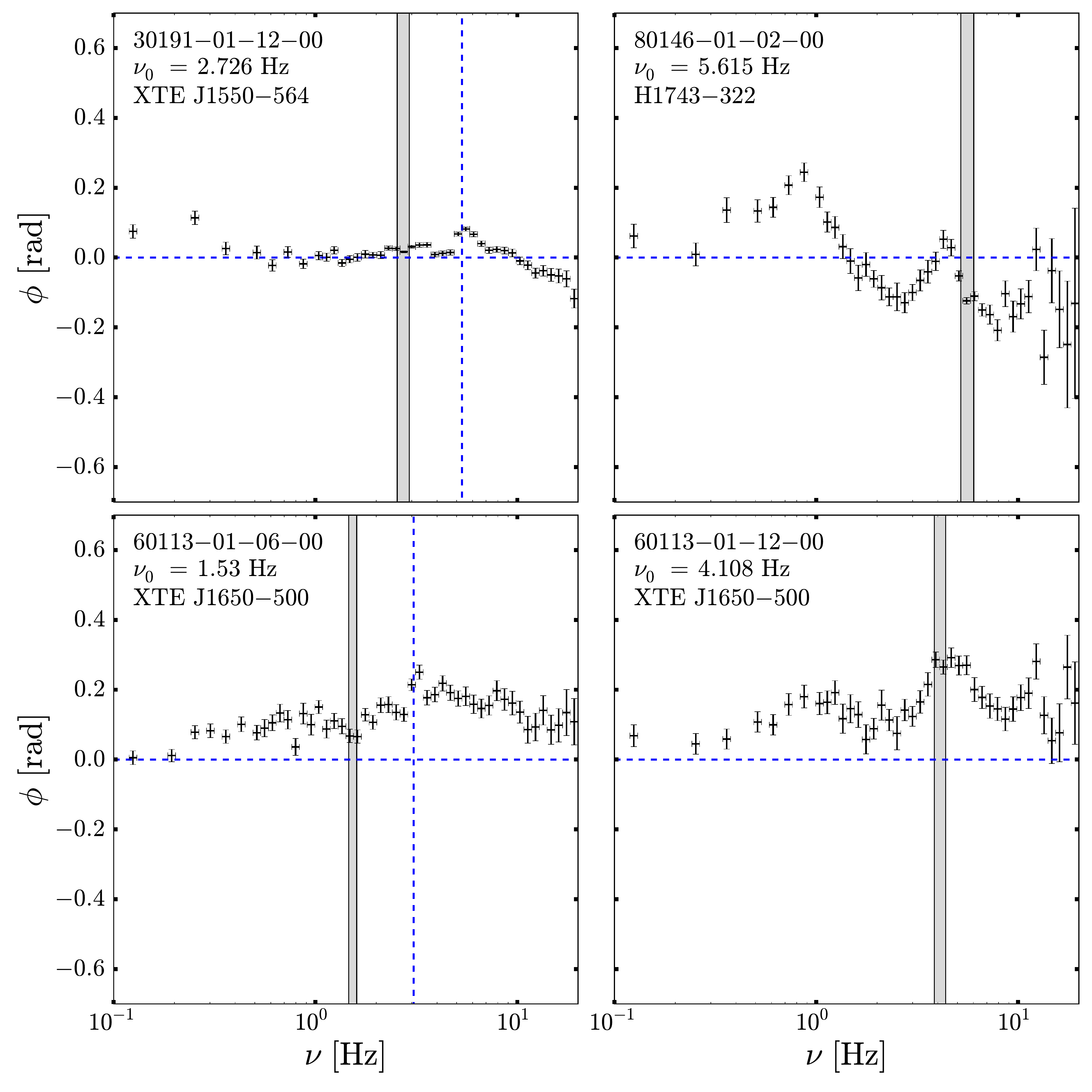}
    \caption{Examples of geometrically binned Type-C QPO lag-frequency spectra. The grey band indicates the range over which the QPO lag is averaged ($\nu_0 \pm \text{FWHM}/2$). The vertical dotted lines indicate (sub)harmonic frequencies.}
    \label{fig:LS_C}
  \end{center}
\end{figure}

In Figures \ref{fig:LS_B} and \ref{fig:LS_C}, we show several examples of logarithmically rebinned lag-frequency spectra of observations showing Type-B and Type-C QPOs, respectively. The QPO fundamental phase lags are averaged over the grey frequency range. The lag at the QPO fundamental is clearly well constrained in this range, regardless of the amount of noise at surrounding frequencies. If present, (sub)harmonic frequencies are indicated with a vertical dashed line. These (sub)harmonic frequencies generally show larger uncentainties in the lags, due to the smaller amplitude of these features. In none of the shown Type-C observations is a (sub)harmonic present in the power spectrum. Generally, the Type-B lag-frequency spectra show larger lag uncertainties than Type-C lag-frequency spectra. 

\begin{figure*}
  \begin{center}
    \includegraphics[width=\textwidth]{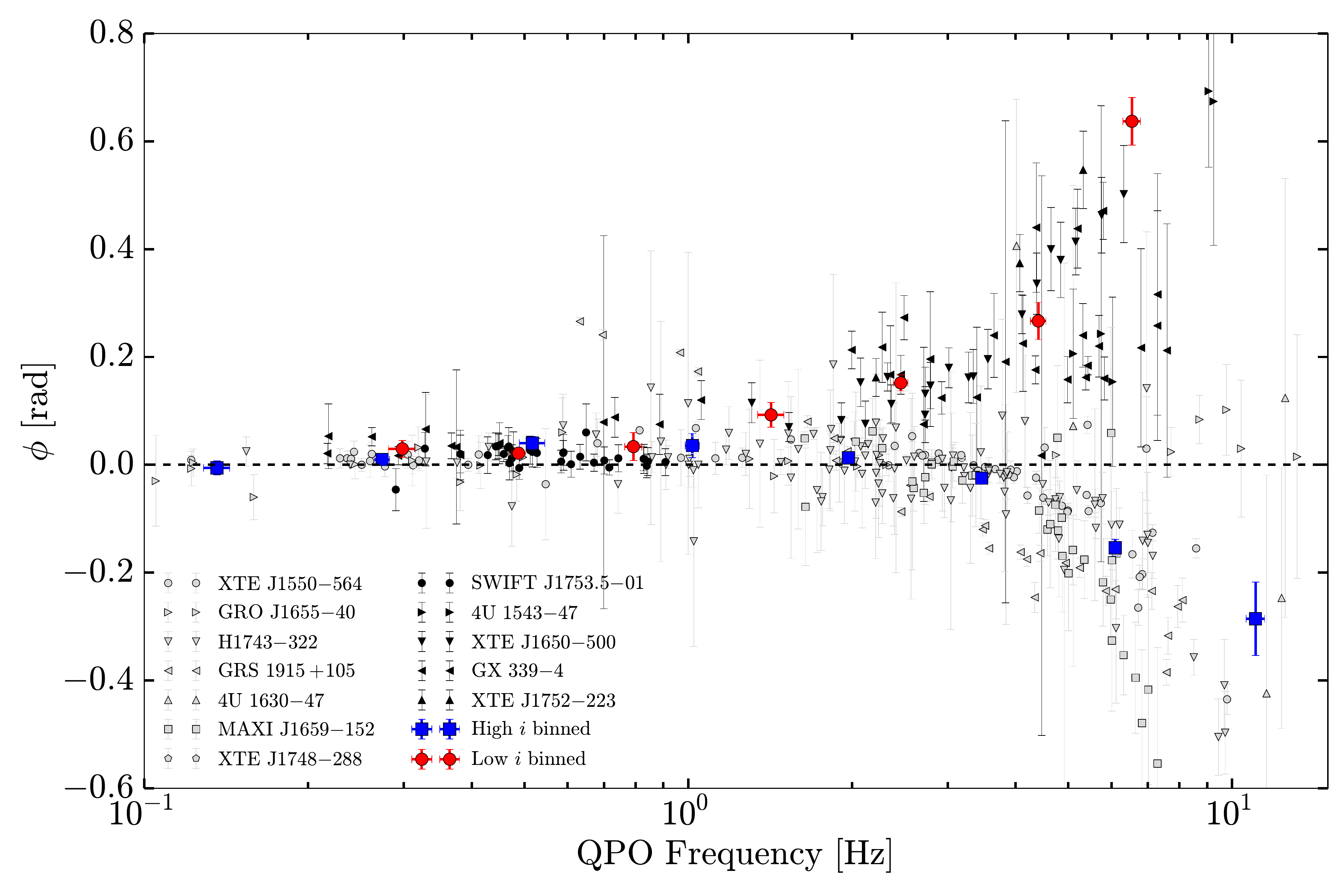}
    \caption{Type-C QPO phase lags as a function of QPO frequency. Black and grey points indicate low and high inclination sources, respectively. Different sources within the low or high inclination sample are shown with different markers. The red and blue points show the average phase lag and QPO frequency in logarithmic frequency bins. The plots for individual sources, including the two sources of unknown inclination, can be found in Appendix A.}
    \label{fig:C_fund}
  \end{center}
\end{figure*}

\subsection{Type-C QPO fundamental lags}

Figure \ref{fig:C_fund} summarizes the main result of this paper: we plot the Type-C QPO fundamental phase lag as a function of QPO frequency for all high (grey points) and low (black points) inclination BHXRBs. For plotting clarity, the high and low inclination lags are binned separately in logarithmic QPO frequency bins and plotted as the blue squares and red circles, respectively. The different markers in the unbinned datapoints represent different sources. All subsequent analysis has been performed on the unbinned data. 

Clear differences are present between the lag behaviour of low and high inclination sources: the dependency of the lags on QPO frequency appears mirrored around a marginally hard average lag: both samples possess a slightly hard lag at low QPO frequencies, at high frequencies high inclination sources turn to soft lags while lags in low inclination sources become harder. In Section 4, we will discuss the peculiar behaviour of GRO J1655-40, which shows no significant changes in lag as a function of QPO frequency, and the lag behaviour of the two sources of unclear inclination (XTE J1752+226 and MAXI J1543-564). In addition, we will also discuss GRS 1915+105, which is the only source showing increasingly hard lags towards low ($<2$ Hz) frequency. Plots of the Type-C QPO lags as a function of QPO frequency for each individual source are available in Appendix A. 

\begin{figure*}
  \begin{center}
    \includegraphics[width=\textwidth]{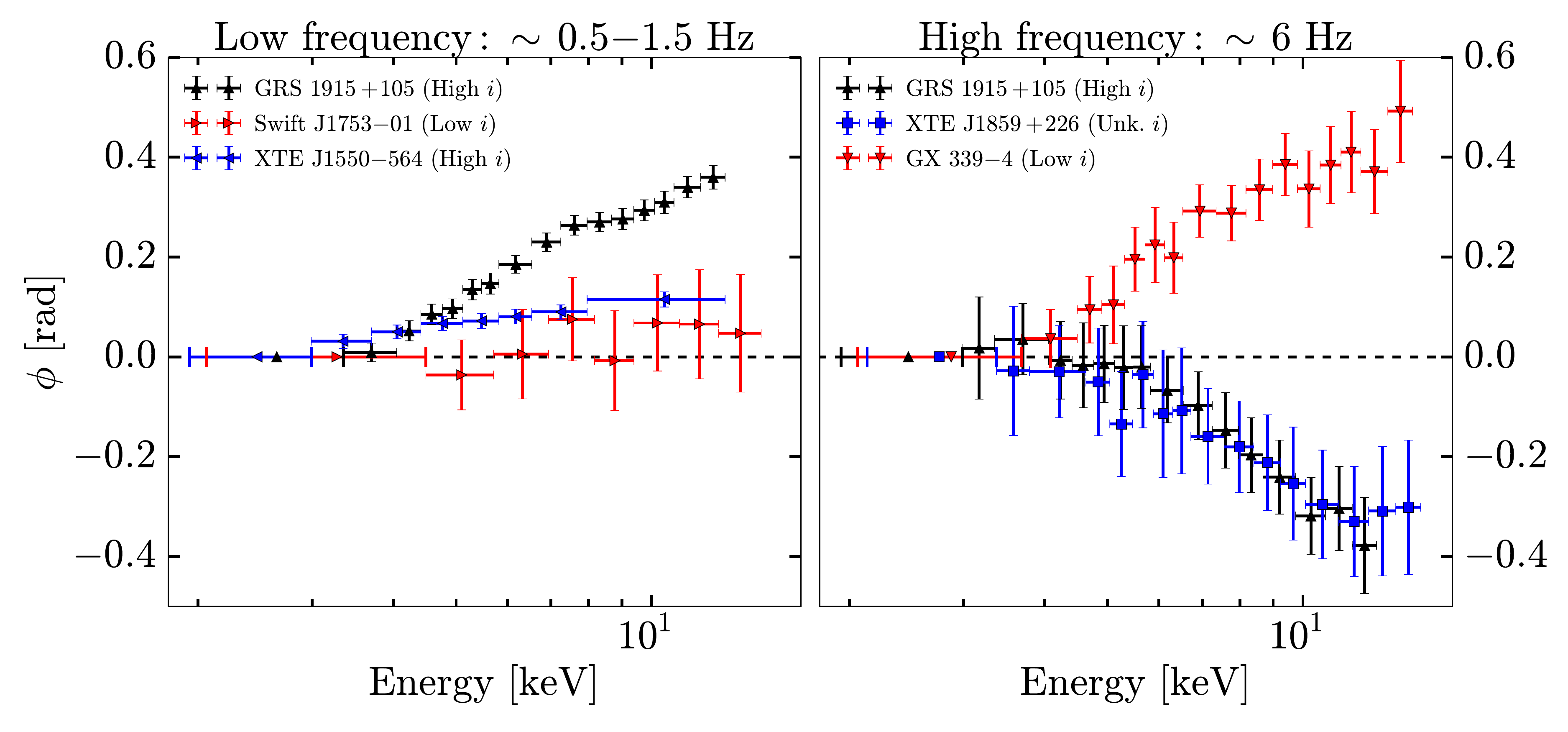}
    \caption{Representative examples of high and low inclination lag-energy spectra for observations with a low ($\sim 0.5$ -- $1.5$ Hz, left) and high ($\sim 6$ Hz, right) Type-C QPO frequency. The sources, RXTE ObsId and inclination are listed in both panels. The reference energy bands are indicated with zero phase lag.}
    \label{fig:LES_C}
  \end{center}
\end{figure*}

To better understand the differences in phase lags between low and high inclination sources, we compare the calculated higher resolution lag-energy spectra for observations with a low ($\sim 0.5$ -- $1.5$ Hz) and high ($\sim 6$ Hz) Type-C QPO frequency. In Figure \ref{fig:LES_C}, we show representative lag-energy spectra for three low and three high QPO frequency observations. The plotted sources include low, high and undetermined inclination sources. For some lag-energy spectra, the uncertainties are relatively large as the QPO is not fully spectrally coherent between all energy bands \citep{vaughan97}. The reference energy band for each observation (generally between $\sim 2$ and $\sim 3-4$ keV, depending on the RXTE data mode) is indicated by the softest energy range with zero phase lag.

At low QPO frequency (left panel), the lag-energy spectra of high and low inclination sources are shaped similarly, as expected based on the similar broad-band lags visible in Figure \ref{fig:C_fund}. GRS 1915+105 however shows distinct behaviour at these low frequencies, as its broad-band QPO lag does not tend to zero but becomes increasingly harder. At high QPO frequency (right panel), the high and low inclination lag-energy spectra are opposite in sign but similar in shape, following a log-linear dependence on energy. The similarity in shape between high and low inclination sources at the same QPO frequency (barring GRS 1915+105 below $2$ Hz) suggests that the underlying mechanism could be the same for hard and soft lags; in that scenario, the inclination is merely resposible for the sign and not the shape of the observed lag and lag-energy spectrum.

Hints of a break, well-known in GRS 1915+105 \citep{pahari13}, are visible as well in the lag-energy spectra of several observations around an energy of $6$--$7$ keV. The similarity of the break-energy to the $6.4$ keV iron K$\alpha$ line suggests that reflection might play a role in the origin of this break. Similarly, its presence could be related to the transition from the disk-dominated to the power-law-dominated region of the spectrum: \citet{stevens16} explicitly show that a similar break in the lag-energy spectrum of the Type-B QPO in GX 339-4 is explained by a difference in lag between the disk blackbody component and the power law component. Furthermore, the monotonic nature of the lag-energy spectra assures that the sign of the broad-band lags does not differs from the sign of the lag between any two narrow bands.

\begin{figure*}
  \begin{center}
    \includegraphics[width=\textwidth]{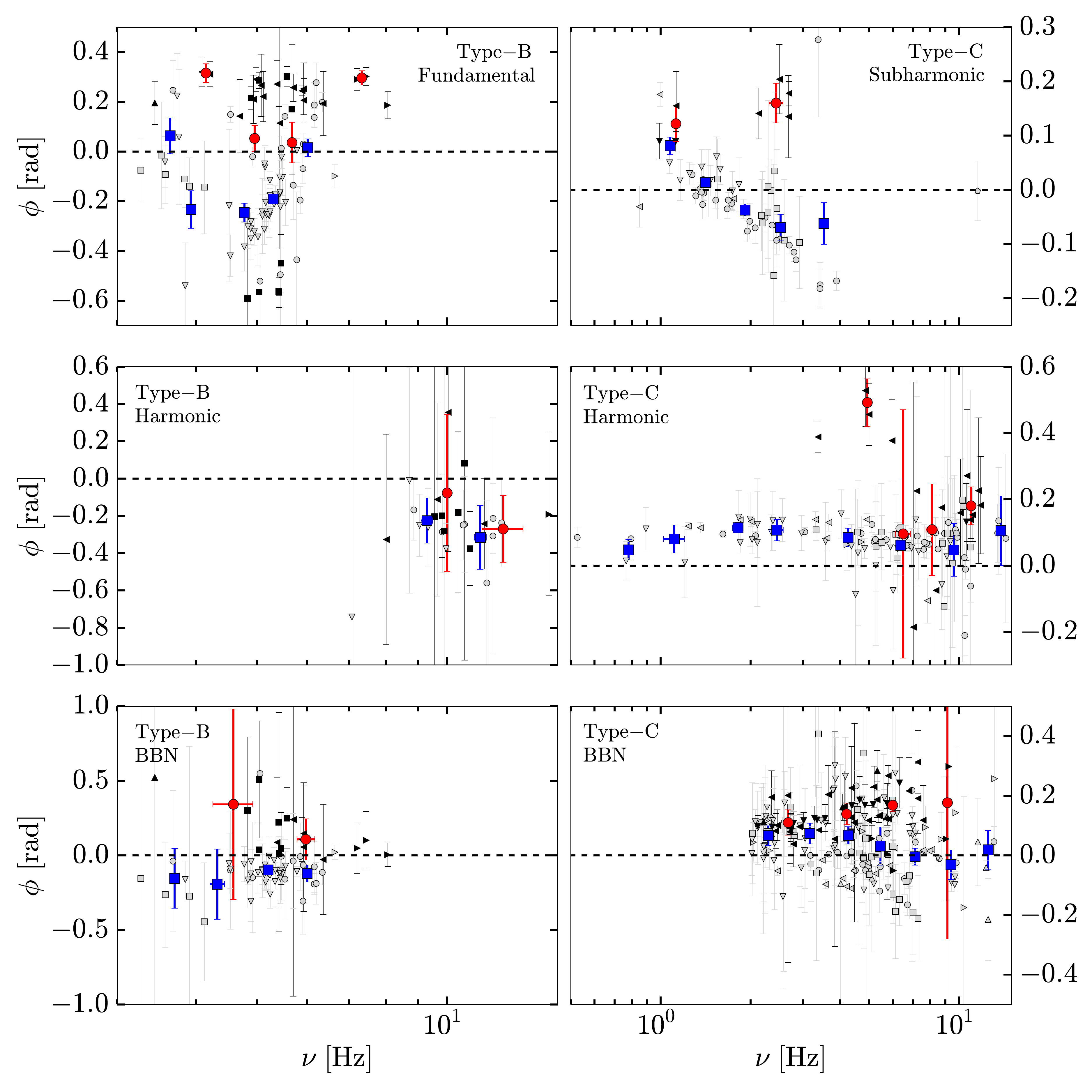}
    \caption{Phase lags averaged over different Fourier frequency ranges for observations containing either Type-B or Type-C QPOs. The left column contains phase lags at the Type-B fundamental, harmonic and BBN. The right column contains phase lags at the Type-C subharmonic, harmonic and BBN. The frequency on the horizontal axes refers to the averaged frequency range, except for the BBN: there it refers to the QPO fundamental frequency. The binning, colors and markers of the datapoints are the same as in Figure \ref{fig:C_fund}. The significance of any differences between the distribution of high and low inclination lags is listed in Table \ref{tab:pvalues}.}
    \label{fig:others}
  \end{center}
\end{figure*}

\subsection{Significance testing}

The Type-C QPO phase lags show a clear difference between low and high inclination sources. However, the number of sources in our sample is quite small. Thus, we want to know how likely it is that the observed differences between low and high inclination could have arisen by chance. In other words, we want to test the null hypothesis that the Type-C QPO phase lags do not depend on the inclination of the source. 

As our test statistic for the difference between phase lags in the two samples, we adopt the absolute difference between the error-weighted average phase lags of the samples:
\begin{equation}
    \label{eq:D}
    D = \left| \sum_{\rm high} \frac{\phi}{\sigma_{\phi}^2} \left/ \sum_{\rm high} \frac{1}{\sigma_{\phi}^2} - \sum_{\rm low} \frac{\phi}{\sigma_{\phi}^2} \right/ \sum_{\rm low} \frac{1}{\sigma_{\phi}^2}\right|
\end{equation}
where \textit{high} and \textit{low} indicate a sum over the high and low inclination observations, respectively. This test statistic offers a simple parameterization of any phase lag differences, without making any assumptions about the dependence on QPO frequency. Hence, it will also be applicable to the lags calculated in other frequency ranges. 

We test our null hypothesis, that the Type-C QPO phase lags do not depend on source inclination, by redistributing the source inclinations and consequently recalculating $D$ $10^4$ times. We redistribute the source inclinations by appointing to each source a new inclination classification, drawn from the set of original classifications ($5\times$low, $7\times$high) without replacement. We then estimate the p-value as the fraction of simulated $D$-values equal to or larger than the $D$-value of the observed data. 

The resulting p-values are listed in Table \ref{tab:pvalues}. As the Type-C QPO lags appear to show the same behaviour in all sources at low QPO frequency, we calculate two p-values: one for the full set of observations, and one with a minimum frequency cutoff of $1$ Hz. At a confidence level of $99\%$, we reject our null hypothesis for the Type-C QPO lags with frequency cutoff. In addition, the Table includes p-values for various other frequency ranges -- these will be discussed in the next sections. 

\begin{table}
 \begin{center}
  \caption{\small{Summary of the significance tests of the inclination dependence of QPO lags. BBN refers to the frequency range $0.5$--$1.5$ Hz, when no QPO/subharmonic is present below $2$ Hz. S and NS refer to the significance/non-significance of the result at a confidence level of $1\%$.}}
  \label{tab:pvalues}
   \begin{tabular}{lllc}
  & $D_{\rm data} $ & P-value & Significance\\
  \hline \hline
  Type-C QPO (no cut) & 0.072 & 0.0408& NS \\
  Type-C QPO ($> 1$ Hz) & 0.236 & 0.0056& S\\
  Type-B QPO  & 0.321 & 0.0600 & NS\\
  \hline
  Type-C harmonic & 0.180 & 0.0490& NS \\ 
  Type-B harmonic & 0.012 & 0.6728& NS\\
  \hline
  Type-C subharmonic & 0.155 & 0.1090& NS \\
  \hline
  Type-C BBN & 0.143 & 0.0016& S\\
  Type-B BBN & 0.169 & 0.0467& NS\\
  \hline
  \end{tabular}
  \end{center}
\end{table}

\subsection{Type-B QPO fundamental lags}

As the Type-C QPO phase lags show a clear dependence on inclination, it is interesting to consider other frequency ranges and types of QPOs. Plots of these additional phase lag results are shown in Figure \ref{fig:others}. All plots are structured and binned in the same manner as Figure 3, including the same markers for the individual sources. Note that XTE J1817-330, which is not included in Figure \ref{fig:C_fund} as there are no Type-C QPO observations of this source in our sample (see Table \ref{tab:sources}), is indicated in Figure \ref{fig:others} by the black squares. The lags in the upper four panels are plotted as a function of their associated frequency (i.e. QPO fundamental, harmonic or subharmonic frequency). The lower two panels (the BBN lags) are plotted against the QPO fundamental frequency, although all BBN lags are calculated in the same frequency range. For consistency, we have binned the lags in all panels in a similar fashion as for Figure \ref{fig:C_fund}. However, in several panels (especially the top left and bottom right), the scatter in lags renders these bin averages physically uninteresting. All statistical analysis and interpretations in this paper are based on the unbinned measurements.

The Type-B fundamental lags are shown in the upper left panel. The lags do not reveal a dependence on inclination as clearly as the Type-C fundamental lags: as is shown in Table \ref{tab:pvalues}, we cannot conclude that the Type-B fundamental lags depend on inclination at a $99\%$ confindence level. However, this might be partially attributable to the smaller number of sources showing Type-B QPOs, and the larger uncertainties on the lags. The observed soft lags are dominated by high inclination sources, namely XTE J1550-564, H1743-322, MAXI J1659-152 and 4U 1630-47. These soft lags are unexpected, as the Type-B QPO classification as defined in \citet{casella05} states that the Type-B fundamental lags are hard. We will discuss this discrepancy further in section \ref{sec:Bdisc}. 

\subsection{Harmonics and BBN}

The remaining panels of Figure \ref{fig:others} show the phase lags at the Type-C harmonic, subharmonic and BBN, and at the Type-B harmonic and BBN. As stated earlier, the BBN lags are calculated over a frequency range from $0.5$--$1.5$ Hz. We do not consider the Type-B subharmonic, as our sample did not contain significant detections of such a feature. The phase lags at the harmonic are predominantly soft and hard for the Type-B and Type-C QPO respectively, as expected based on the QPO Type-Characteristics listed in \citet{casella05}. Interestingly, the Type-C subharmonic phase lags show both hard and soft lags, with all soft lags observed in high inclination sources. Lastly, the BBN phase lags cover a wide range in values for both QPO types, without an apparent dependence on the QPO frequency. 

Interestingly, as Table \ref{tab:pvalues} shows, the BBN lags of the Type-C QPO depends significantly on inclination at a $99\%$ confidence level. Although differences between the high and low inclination sample appear to be present in some other cases, especially at the Type-C subharmonic, the number of low inclination sources is too low to find a significant inclination dependence. It is however interesting to note that for the Type-C subharmonic, only high inclination sources show soft lags. Additionally, it is notable that the Type-C harmonic lags are remarkably constant for high inclination sources, while the lags in low inclination sources do show variations between sources. 

\begin{figure*}
  \begin{center}
    \includegraphics[width=\textwidth]{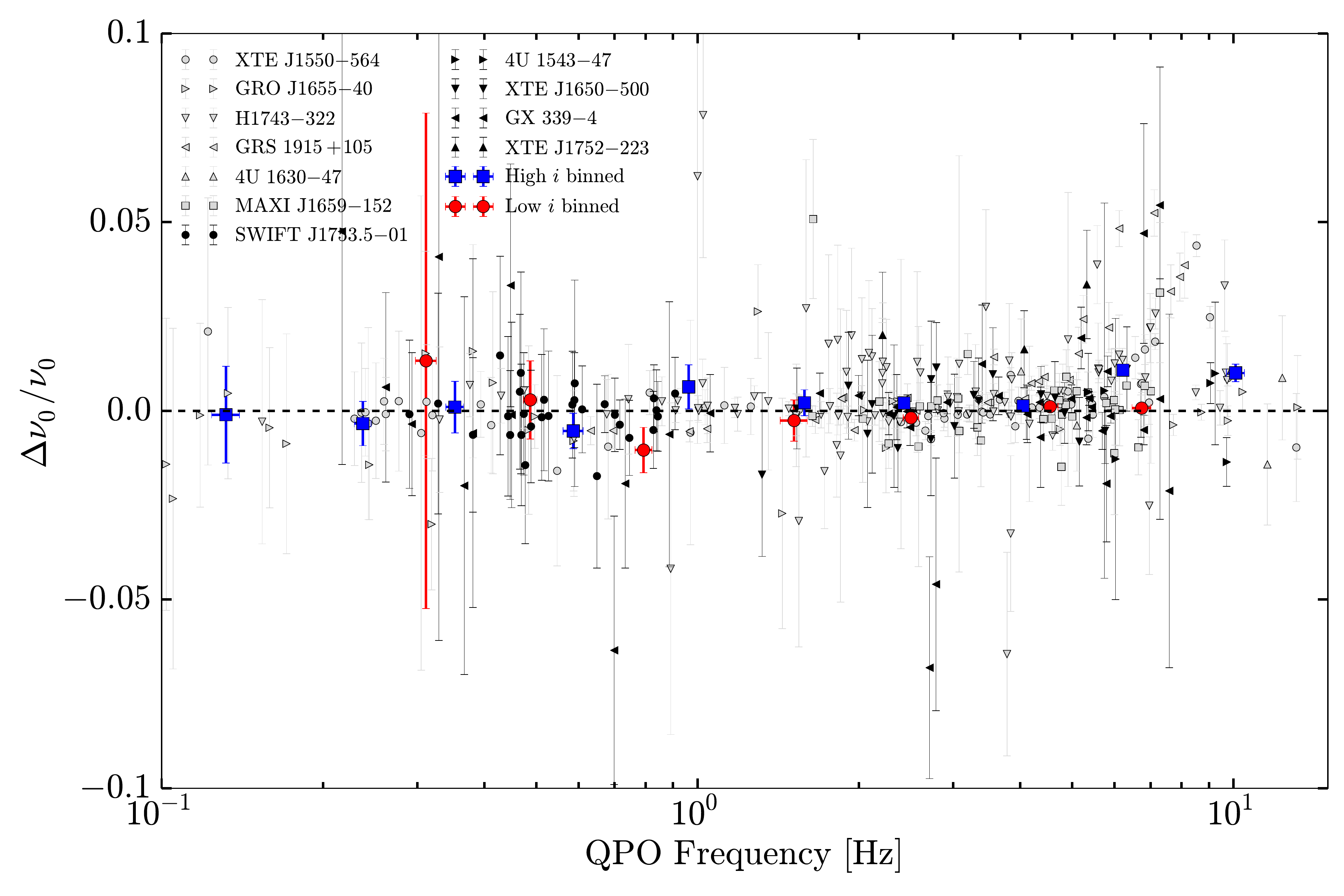}
    \caption{The relative frequency difference between the Type-C fundamental centroid in the hard and soft band $\Delta \nu_0/\nu_0$, as a function of QPO frequency in the full band $\nu_0$. The binning, markers and colors of the datapoints are the same as in Figure \ref{fig:C_fund}.}
    \label{fig:df_all}
  \end{center}
\end{figure*}

\subsection{Energy-dependent frequency differences of the Type-C Fundamental}

In addition to the inclination dependence of the average phase lag, we also investigate the short-timescale evolution of the phase lag by measuring differences in Type-C fundamental frequency between the soft and hard band. We plot these frequency differences, divided by the QPO frequency itself, as a function of Type-C QPO frequency in Figure \ref{fig:df_all}, in a similar fashion to Figures \ref{fig:C_fund} and \ref{fig:others}. Again, we bin low and high inclination sources seperately for clarity, while plotting the unbinned data in black and grey, respectively. Low inclination source XTE J1748-288 is not included, as the error on the fitted QPO centroid frequency is too large to accurately determine frequency differences. 

There is no immediately clear difference between high and low inclination sources, contrary to the average Type-C QPO phase lags. At low QPO frequency, few significantly non-zero frequency differences are present, as was known from earlier investigations \citep{qu10,li13,li13b}. At higher QPO frequencies, non-zero frequency differences become more prevalent, but due to large uncertainties, these are rarely significant. Hence, no clear difference can be found when grouping the sources as low and high inclination. 

\begin{figure}
  \begin{center}
    \includegraphics[width=\columnwidth]{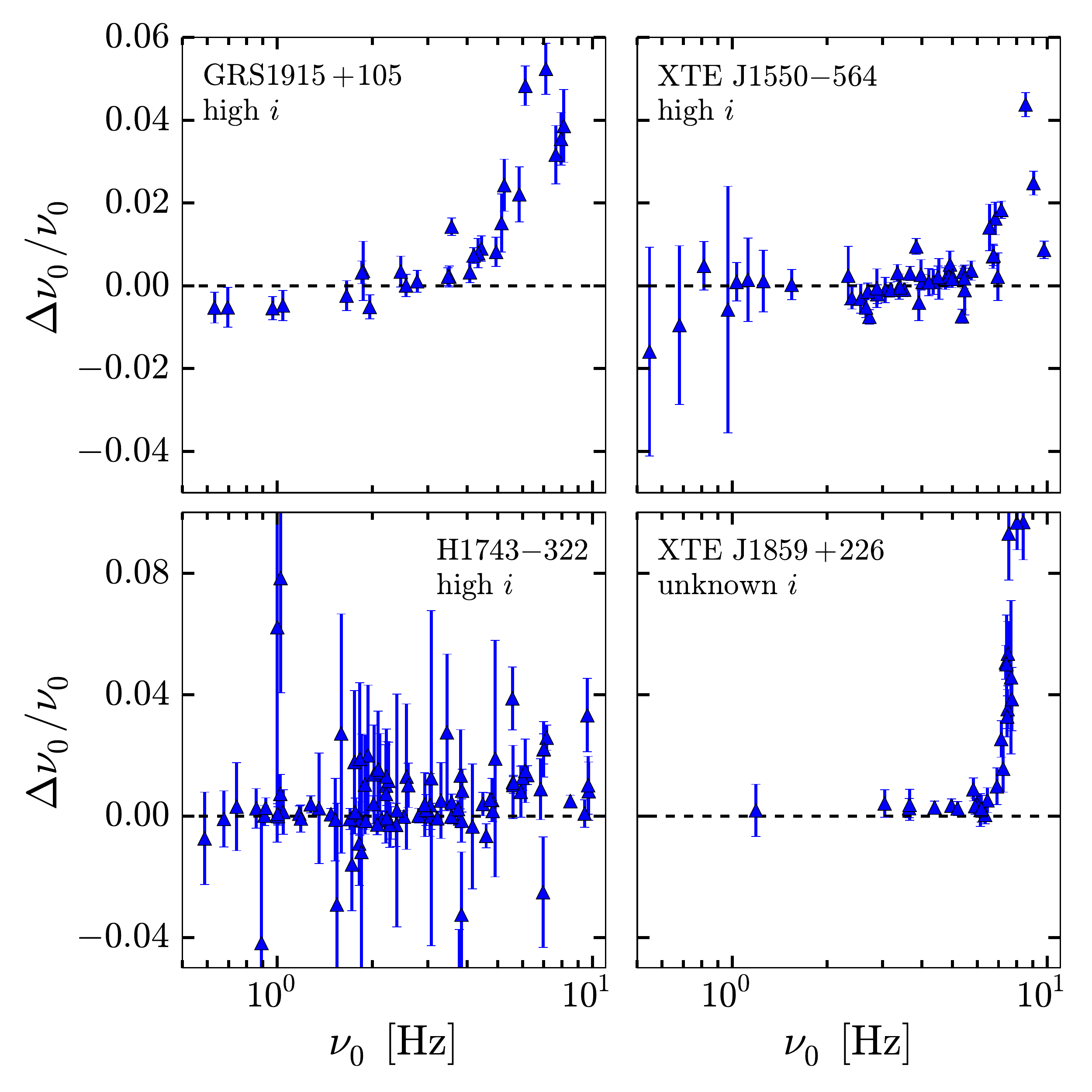}
    \caption{Relative frequency difference between the Type-C centroid in the hard and soft band $\Delta \nu_0/\nu_0$, as a function of QPO frequency in the full band $\nu_0$, for four sources: GRS 1915+105, XTE J1550-564, H1743-322 and XTE J1859+226. Only these four sources show significant non-zero frequency differences. Similar plots for all sources indiviually can be found in Appendix A.}
    \label{fig:df_inc}
  \end{center}
\end{figure}

In order to investigate the presence of QPO frequency differences in more detail, we consider them for all individual sources, independent of inclination. For each source, we use a $\chi^2$-test to test whether the distribution of frequency differences is consistent with zero within the $99\%$ confidence limit. This returns four sources with a significant relative frequency difference, which are shown in Figure \ref{fig:df_inc}. The three high inclination sources (GRS 1915+105, XTE J1550-564 and H1743-322) were already known to show frequency differences \citep[see respectively][]{qu10, li13, li13b}. The fourth source, with unknown inclination (XTE J1859+226), also shows clear non-zero frequency differences at high QPO frequencies. In addition, XTE J1752-223 (low inclination) shows marginally significant frequency differences ($p=0.02$). However, we analysed only three Type-C observations for this source, so the significance of this result should be regarded with caution. No other low inclination sources show hints of significant frequency differences. In appendix A, we plot the frequency difference as a function of QPO frequency for each individual source. 

\section{Discussion}
\label{sec:disc}

In this paper, we present the results of a systematic, model-independent analysis of the inclination dependence of QPO phase lags and frequency differences. We report the following two main results:
\begin{enumerate}
    \item[(1)] the phase lag at the Type-C fundamental QPO frequency depends significantly on source inclination, both in sign and relation to QPO frequency, and
    \item[(2)] four sources (three high inclination and one undetermined inclination) show significant frequency differences in the Type-C QPO.
\end{enumerate} 
In this section, we will discuss and interpret all Type-C and Type-B results, the absence and presence of frequency differences, and the behaviour of several individual sources. In addition, we will also discuss the adopted inclination estimates.

\subsection{Type-C QPO lags}
\label{sec:typeC}

For the Type-C QPOs, we conclude that phase lags at the fundamental and the BBN depend significantly on inclination. This finding is consistent with the result in \citet{motta15} that the amplitude of both the QPO fundamental and the BBN depend on inclination. Furthermore, we find that the phase lags at the harmonic are, with a handful of exceptions, hard - high inclination sources even show a remarkably constant lag as a function of QPO frequency, as was known for GRS 1915+105 \citep{pahari13}. Lastly, it is important to note that for the QPO fundamental, inclination does not suppress or magnify the value of the observed phase lags. Instead, the signs are completely flipped, while the amplitudes remain comparable. Any physical interpretation of these results should be able to explain this behaviour. 

Calculating the cross spectrum between two lightcurves does not allow for the direct distinction between QPO and BBN lags. In the example Type-C QPO lag-frequency spectra (Figure \ref{fig:LS_C}), lag features are visible at other frequencies than only the QPO frequency; for instance, in the top-right lag-frequency spectrum (H1743-322), the QPO lag is located on the slope of a larger lag feature at lower frequency. In the remainder of this discussion, we will assume that the lag at the QPO frequency is dominated by the QPO itself. However, we note that the BBN might also influence the measured lags. The remainder of this discussion of Type-C QPO lags is divided into two parts, focussing on geometric and intrinsic models, respectively. Given the possible influence of BBN lags, and lack of detailed quantitative lag predictions in current QPO models, we merely discuss the general differences between the high and low inclination sources, such as Type-C QPO lag sign.

\subsubsection{Geometric origin: Lense-Thirring precession}

In this section, we consider a physical explanation in the context of the Lense-Thirring precession model, wherein the inclination dependence follows from differences in dominant general-relativistic effects on the observed flux along different lines of sight. In this interpretation, the lags themselves (independent of inclination) originate from spectral pivoting during each precession cycle. \citet{ingram15}, \citet{ingram16} and Stevens \& Uttley (in prep.)  use phase resolved spectroscopy of GRS 1915+105, H1743-322 and GX 339-4 respectively to show that the power law spectral index $\Gamma$ oscillates as a function of Type-C QPO phase. Although QPO phases cannot be mapped directly onto precession phases, this implies that the hardness changes during a precession cycle. If the $\Gamma$ variation peaks after the QPO flux peaks, this causes a soft lag at the QPO frequency, and vice versa. Making the assumption that the spectral pivoting as a function of precession phase is independent of source inclination, both the sign and size of the observed lag will simply depend on the QPO flux as a function of precession phase. 

\begin{figure}
  \begin{center}
    \includegraphics[width=\columnwidth]{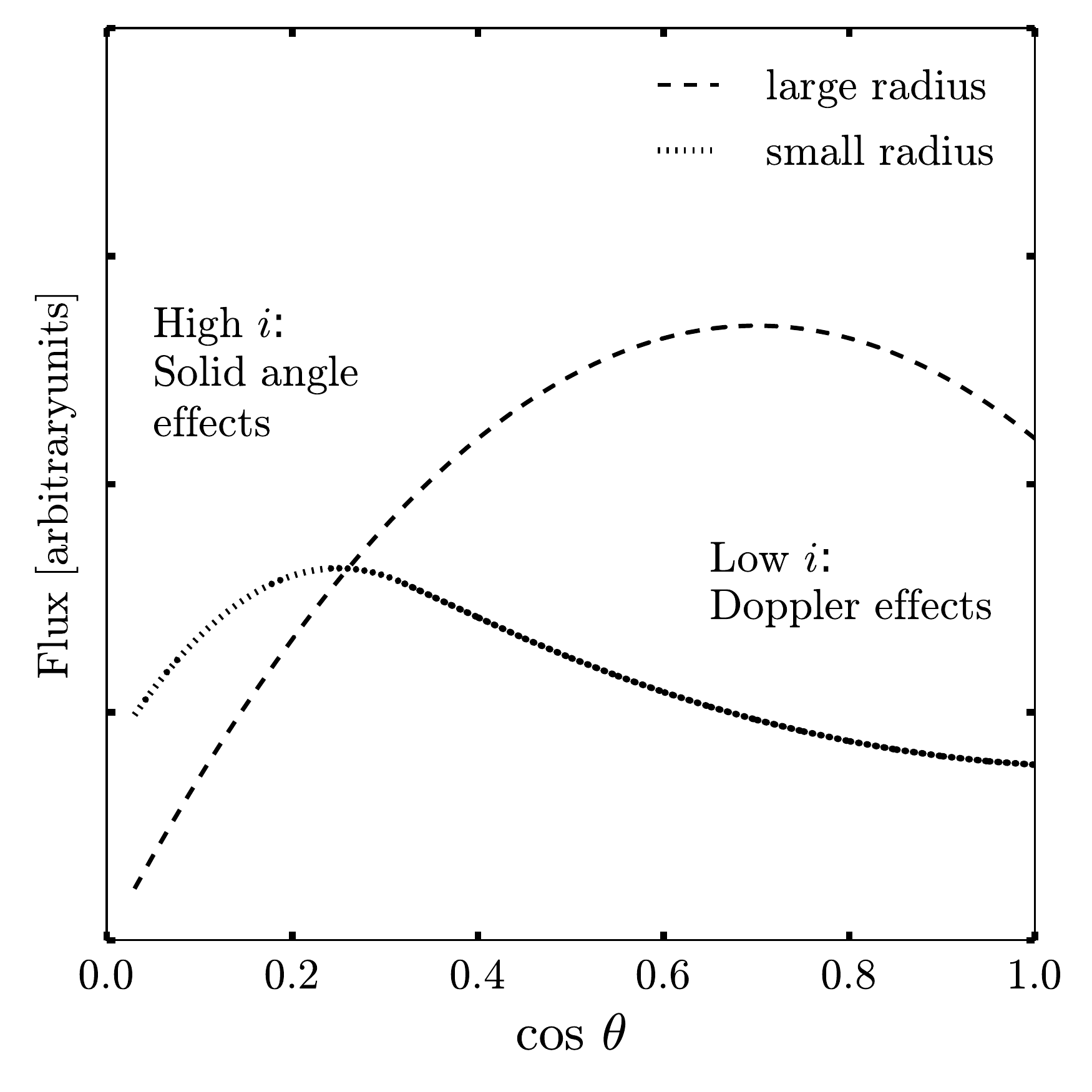}
    \caption{Sketch of Figure 3 from \citet{veledina13}: the observed flux of the simulated precessing ring as a function of the cosine of the viewing angle $\theta$. Large $\theta$, shown on the left, corresponds to a high inclination (i.e. edge on) of the ring. The flux, fully modeled by \citet{veledina13}, is sketched for a small and large ring.} 
    \label{fig:valedina}
  \end{center}
\end{figure}

\citet{veledina13} model a precessing inner flow as a precessing Comptonizing ring, taking relativistic effects into account. In Figure \ref{fig:valedina}, we show a sketch of their Figure 3: the dependence of the observed flux on inclination for emission from different radii. The angle $\theta$ is the instantaneous inclination angle of the flow as seen by an observer -- small $\theta$ corresponds to a face on precessing flow. The angle $\theta$ changes systematically during a precession cycle. Figure \ref{fig:valedina} shows that, for small radii, this has the opposite effect on the observed flux for low and high inclination sources: high inclination sources become brighter as $\theta$ decreases ($\cos \theta$ increases), while for low inclination sources this effect is opposite. This differences arises due to differences in dominant relativistic processes: Doppler effects at low inclination versus solid angle effects at high inclination\footnote{Note that the limb darkening law also plays role -- however, as it does not depend on radius, we do not consider it in this interpretation.}.

As mentioned earlier, the dependence of flux on precession phase could set both the sign and size of the observed QPO lag. Hence, this difference in dominant relativistic effects between inclinations could result in the observed inclination dependence of the Type-C QPO phase lag. As is visible in Figure \ref{fig:valedina}, this effect might become less important at larger radii, as the curve's maximum shifts to high values of $\cos \theta$ and differences in relativistic effects become smaller (i.e. solid angle effects dominate for nearly all values of $\cos \theta$). This might explain the similarity in the Type-C QPO lags at low QPO frequencies, which correspond to larger radii. 

The origin of the QPO phase lags and their inclination dependence can be probed more accurately using detailed lag-energy spectra. At low QPO frequency, high and low inclination sources display similar lag-energy spectra, just as they show similar average lags. At high QPO frequency, low and high inclination sources still show similarly shaped lag-energy spectra, but of opposite sign. This similarity in shape is consistent with our proposed interpretation of the inclination dependence of the broad-band lags: the lags themselves always originate from the same mechanism (the pivoting power law), while the sign is set by only the inclination through relativistic processes. In this scenario, it is expected that the sign of the lag-energy spectra would change, but not the shape. At low QPO frequencies, GRS 1915+105 is an exception: its lag-energy spectrum is not flat but similar to low-inclination high-QPO-frequency lag-energy spectra, suggesting an additional source of lags. We will discuss this behaviour in Section \ref{sec:grs}.

Alternatively, Lense-Thirring precession could cause Type-C QPOs through diskoseismic models. In such models, QPOs arise from the oscillation modes of a thin accretion disk, excited in a general-relativistic potential \citep{kato80, wagoner99}. Diskoseismic models are mostly used to explain the presence of HFQPOs. However, the lower-frequency Type-C QPOs could arise from corrugation modes: vertical oscillation of the disk , trapped in between its inner edge and the inner vertical resonance. The frequency of these oscillations matches the Lense-Thirring frequency at this inner vertical resonance \citep[e.g.][]{tsang13}.

The interpretation of the inclination-dependent Type-C QPO lags, discussed in this section, consists of two steps: the observed variation in spectral index $\Gamma$ within QPO cycles, and the trade-off between relativistic effects in a vertically precessing ring viewed at different inclinations. The former is a model-independent phenomenon, while the vertical precession is present in diskoseismic models through the corrugation modes. However, one important difference should be noted: in diskoseismic models, the QPO originates from the disk instead of the inner flow, while the QPO is strongest in the spectral components originating in the inner flow \citep{sobolewska06, alexsson13}. Hence, the precession-model by \citep{ingram09} appears to match the Type-C QPO properties more generally than diskoseismic models.

The phase lags at the Type-C harmonic show a completely different behaviour than those at the fundamental. The difference between high and low inclination sources is not significant at a $99\%$ confidence level, as the sample contains a relatively small number of low inclination observations. More importantly, the lags are clearly hard, with little dependence on QPO frequency. Thus, contrary to the fundamental, in our sample the inclination does not have an influence on the sign of the phase lags at the harmonic. This points to a difference in the exact physical mechanism creating the harmonic signal.

Using frequency-resolved spectroscopy, \citet{axelsson14} show that the spectrum of the harmonic is softer than both the spectrum of the funamental and the time-averaged spectrum. This indicates that the origin of the harmonic might be related to the thermal reprocessing and reflection of hard photons by the disk. Furthermore, \citet{ingram16} show that in H1743-322, several spectral parameters vary during each QPO cycle. However, these parameters show strong harmonic features (i.e. a second minimum and maximum) within each QPO cycle. This effect is interpreted as resulting from a precessing inner flow, which is inclined with respect to the disk and thus illuminates the disk on two sides: both the front and the back. In such an interpretation, this behaviour could also give rise to a harmonic at twice the QPO frequency. These results on the possible role of reflection in the harmonic tie in to our lag results, since the inclination of the disk with respect to the precessing inner flow is independent of the position of the observer. Hence, the position of the observer will not have a direct influence on the observed sign of the phase lag, as is the case with the fundamental. Instead, the lags might be set by the time difference between illuminating the receding and approaching side of the disk. In that scenario, red and blueshifting of the reflected harmonic could cause a lag that is independent of the observer's position. 

\subsubsection{Intrinsic origin}

It is less intuitive to account for an inclination dependence of QPO properties in intrinsic models, such as accretion rate fluctuations in the disk \citep{tagger99,cabanac10} or standing shocks in the accretion flow \citep{chakrabarti93}. In such intrinsic models, although no quantitative lag predictions are made, the QPO lags could originate from different mechanisms. For instance, Comptonization, where photons from the disk are upscattered in the corona, creates lags compared to photons that are not upscattered. Alternatively, the propagation of accretion rate oscillations to regions of different temperatures can introduce energy-dependent lags. As we will detail below, these two mechanisms are not able to qualitatively explain either the size nor the inclination dependence of the Type-C QPO lags.

At a Type-C QPO frequency of approximately $5$ Hz, we typically observe phase lags around $\pm 0.5$ rad. This corresponds to hard and soft time lags of approximately $16$ ms. For such time lags to originate from Comptonization, the corresponding Comptonizing regions need to be unrealistically large: for a typical $10$ Solar mass black hole, a light travel time of $16$ ms corresponds to a distance of $\sim 160$ $r_g$ (although the actual size of the corona will be smaller when the optical depth is high). Note that, as unscattered \'and scattered photons are present in both the hard and soft energy band, this distance is merely an absolute lower limit, while the actual distance is much larger.

Furthermore, to explain the inclination-dependent switch from soft (high inclination) to hard (low inclination) lags, the Comptonizing process needs to depend on inclination: at low inclination, inverse-Compton scattering should increase the energy of lagging photons, resulting in hard lags. Alternatively, at high inclination, regular Compton scattering, decreasing the photon energies, should take over and cause soft lags. While a coronal geometry resulting in such effects might be conceivable, this scenario would have large effects on the spectral appearance: high inclination sources should show systematically softer spectra. In fact, the opposite is the case: \citet{heil15} show that high inclination sources show harder spectra.

A similar argument holds for the propagation of oscillations in the disk. To account for an inclination-dependent lag in such models, the observed temperature gradient of the disk should depend on inclination: switching from an observed increasing temperature towards the black hole at low inclination (resulting in hard lags) to an observed decreasing temperature at high inclination (resulting in soft lags). However, since the inner regions should contribute more to the luminosity, this scenario would also result in systematically softer spectra at high inclination, opposite to the observed behaviour in \citet{heil15}. Thus, intrinsic models are not able to qualitatively explain the observed relation between Type-C QPO lags and inclination.

Recently, \citet{dutta16} have suggested that the different hard and soft lags seen respectively in GX~339-4 and XTE~J1550-564 for high QPO frequencies could be linked to their different inclinations.  Their model invokes different relative contributions of inverse Compton scattering and lagging of disk and outer hot flow soft photons by light bending.  However, to reproduce the observed lags, their model assumes that the disk is truncated at around 200~gravitational radii or larger: this large truncation radius is inconsistent with the strong, high temperature (up to 0.8~keV) disk blackbody component seen in the spectrum for the same QPO frequencies \citep[e.g.][]{sobczak00, remillard02}, which instead suggests a much smaller truncation radius, consistent with that expected from precession.

\subsection{Inclination estimates}
\label{sec:misalignments}

In our analysis, we have classified sources based on binary-orbit inclination. In this procedure, we have classified all sources showing absorption dips as high inclination. These absorption dips, occuring periodically for $10$--$30$\% of the orbit, are caused by a bulge at the interaction point between the disk and the companion star \citep{white82}. Although dips were reported in low-inclination source 4U 1543-47 \citep{park04}, these are of a different nature: they occur on much shorter timescales than the orbital period \citep{orosz98}. Similarly, ultra-luminous X-ray sources, such as NGC 5408 X-1, are known to show dips \citep{pasham13}. However, as \citet{grise13} show, these dips are not purely periodic, leaving it unclear whether they are linked to binary motion.

An important difficulty in the application of binary-orbit inclinations is the possibility of misalignment between the binary orbit and the inner disk. The QPO is thought to originate from the inner disk, and recent results strongly suggest that the Type-C QPO originates from precession around the black-hole spin axis \citep{ingram15, ingram16, motta15}. For that reason, it is important to compare our binary-orbit inclination estimates with inner-disk inclination estimates. As such, we can check whether the binary orbit can indeed be applied as an accurate tracer of the inner-disk inclination.

Two methods of estimating this inner-disk inclination exist: measuring the inclination of the jet, and estimating the inclination of the inner disk through the iron K$\alpha$ line in the reflected spectrum\footnote{The iron K$\alpha$ line originates from the optically thick, geometrically thin accretion disk. Hence, the inclination might still not be exactly equal to the inner-disk inclination. However, it should at least be closer in alignment than the binary orbit}. Of the fifteen sources in our sample, nine have such estimates of the inner disk inclination. Table \ref{tab:sources_innerdisk} lists these nine sources, including the estimated inner-disk inclinations and the method of obtaining these inclinations. It is important to note that all sources for which the inner-disk inclination is accuretely determined, remain within the same inclination classification. For Swift J1753.5-01, the inner-disk inclination estimates appear to depend heavily on the spectral model, spanning a large range in fitted inclinations \citep{hiemstra09}. Hence, we do not reclassify it as either high or low inclination. However, this source shows no Type-C QPOs above $1$ Hz, where the dependence of the Type-C QPO lags on binary-orbit inclination is observed.

It is not surprising that all sources with accurate inner-disk inclination estimates remain in the same inclination classification; as stated in Section \ref{sec:2}, the broad classification into merely \textit{low} and \textit{high} inclination allows for a misalignment between the inner disk and the binary orbit. Although the sources with accurate estimates for both inclinations do show slight misalignments, these differences are not large enough to change any classifications. 

We have repeated our statistical analysis of the dependence of the Type-C QPO lag on inclination, using the new inclination classification based on inner-disk inclinations (Table \ref{tab:sources_innerdisk}). This yields a slightly lower significance for observation with QPO frequencies above $1$ Hz, with $p=0.0157$. However, this slight decrease in significance is expected based on the decrease in number of sources and observations: none of the sources with observations above $1$ Hz changes classification, but the number of sources decreases by one third from twelve to eight. In other words, the lower significance appears not to follow from a lack of dependence on inclination but from a lack of data.

\begin{table}
 \begin{center}
  \caption{\small{Summary of the source sample based on inner-disk inclination estimates. The methods listed are \textit{R.R.}: relativistic-reflection spectrum modeling, and \textit{Jet}: measuring the jet inclination. Ranges in inclination indicate inclination estimates based on different spectral models. A $\sim$ indicates that different spectral models result in inclinations close to the shown value. References for the inclination measures (2$^{\rm nd}$ column): [a] \citet{morningstar14}, [b] \citet{miller02}, [c] \citet{miller04a, miller04b}, [d] \citet{reis11}, [e] \citet{hiemstra09}, [f] \citet{steiner12b}, [g] \citet{hjellming95}, [h] \citet{steiner12}, [i] \citet{mirabel94}}}
  \label{tab:sources_innerdisk}
   \begin{tabular}{lcccc}
  \hline
  Source & Inner-disk $i$ [$^{\rm o}$] & method & Sample & Ref. \\
  \hline \hline
  4U 1543-47 & $\sim 30$ & R.R. & Low & [a]\\
  XTE J1650-500 & $\sim 45$ & R.R. & Low & [b] \\
  GX 339-4 & $\lesssim 45$ & R.R. & Low & [c] \\
  XTE J1752-223 & $16-28$ & R.R. & Low & [d] \\ 
  \hline
  Swift J1753.5-01 & $0.9 - 90$ & R.R. & - & [e] \\
  \hline
  XTE J1550-564 & $70.8^{+7.3}_{-4.5}$ & Jet & High & [f] \\
  GRO J1655-40 & $85 \pm 2$ & Jet & High & [g] \\
  H1743-322 & $75 \pm 3$ & Jet & High & [h] \\
  GRS 1915+105 & $70 \pm 2$ & Jet & High & [i] \\
  \hline
  \end{tabular}
  \end{center}
\end{table}

\subsection{Type-B QPO lags}
\label{sec:Bdisc}

Despite apparent differences in Figure \ref{fig:others}, the Type-B fundamental phase lags do not show a dependence on source inclination at a $99\%$ confidence level. Furthermore, these lags show a different dependence on QPO frequency than the Type-C lags. This adds to the evidence that the Type-B QPO arises from a different mechanism, \citep{motta12,fender09}, such as the precessing base of a jet \citep{motta15,stevens16}. However, it is important to note that the data quality for the Type-B QPOs is less than for the Type-C QPOs: the uncertainties are larger and the number of observations and sources is smaller. If the Type-B QPO indeed originates from a jet-based mechanism, an inclination dependence in the phase lags might be expected. It is possible that the data quality in our analysis is simply too low to significantly detect such a dependence. 

The soft lags we observe in several high inclination sources are not expected based on the Type-B QPO classification as defined by \citet{wijnands99} and \citet{casella05}, which describes only hard lags in Type-B QPOs. We observe these unexpected lags in XTE J1550-564, H1743-322, 4U 1630-47 and MAXI J1659-152. From these sources, only XTE J1550-564 was included in the analysis of the original definition of QPO types. However, our sample of Type-B observations in XTE J1550-564 is more extensive than the original sample. Furthermore, \citet{wijnands99} define the fundamental QPO as the $\sim 3$ Hz peak, while we define this as the subharmonic. Thus, it appears to be too simple to state that all Type-B QPO fundamentals have hard lags. Based on our results, these lags could be both hard and soft. 

\subsection{Broad-band noise}

The phase lags of the BBN associated with the Type-C QPO (in the $0.5$--$1.5$ Hz range) show a significant inclination dependence at the 99\% confidence level, while those associated with the Type-B QPO do not. In both cases, the soft lags appear to be dominated by the high inclination sources. For Type-B QPO observations, marginally hard lags are seen in low inclination sources. For Type-C QPO observations however, hard lags are observed in both high and low inclination sources. For both the Type-B and Type-C observations, the BBN lag does not show any clear dependence on the QPO frequency (as is clearly present in for instance the Type-C QPO lag).

Existing results on the inclination dependence of the BBN are not in full agreement: \citet{heil15} find that the BBN, unlike the QPO, does not show a dependence on inclination. \citet{motta15} however show that the amplitude of the BBN associated with the Type-C QPO does depend on inclination. In comparing these results with ours it is important to note that \citet{heil15} analyse observations regardless of the presence of the QPO. Our results on the other hand are derived from observations selected on QPO type. The bias introduced by this selection might partially explain the inclination dependence observed for the Type-C BBN: if the physical origin of the QPO and BBN are related, the presence of a QPO might result in inclination dependent properties in the BBN. Alternatively, the BBN lags might simply be contaminated by QPO lags, despite our cut on QPO and subharmonic frequency. 

Furthermore, the $0.5$--$1.5$ Hz range is quite arbitrary, chosen merely for its relatively large BBN amplitude. Thus, we have repeated the analysis on the BBN lags averaged over the range $0.1$--$0.5$ Hz in the same observations (i.e. those with no QPO or subharmonic below $2$ Hz). As this range covers lower frequencies, possible contamination by the QPO lags is less likely. In this new range, no significant inclination dependence is present in both QPO types (Type-C: $p=0.0729$, Type-B: $p=0.73$), despite similar data quality. Thus, as the inclination dependence of Type-C BBN lags between $0.5$--$1.5$ Hz appears to be closely related to the arbitrary frequency range, we do not draw general conclusions about the nature of the BBN associated with Type-C QPOs. 

\subsection{Individual sources}

In our analysis, we have grouped sources into high, low or undetermined inclination and assumed that any significant differences in lag behaviour originate from differences in inclination. In this section, we will discuss two sets of individual sources: sources of undetermined inclination (XTE J1859+226 and MAXI J1543-564) and sources that do not follow the standard trend in Type-C fundamental lags as function QPO frequency (GRO J1655-40 and GRS 1915+105). 

\subsubsection{XTE J1859+226 and MAXI J1543-564}
\label{sec:undetermined}

\begin{figure}
  \begin{center}
    \includegraphics[width=\columnwidth]{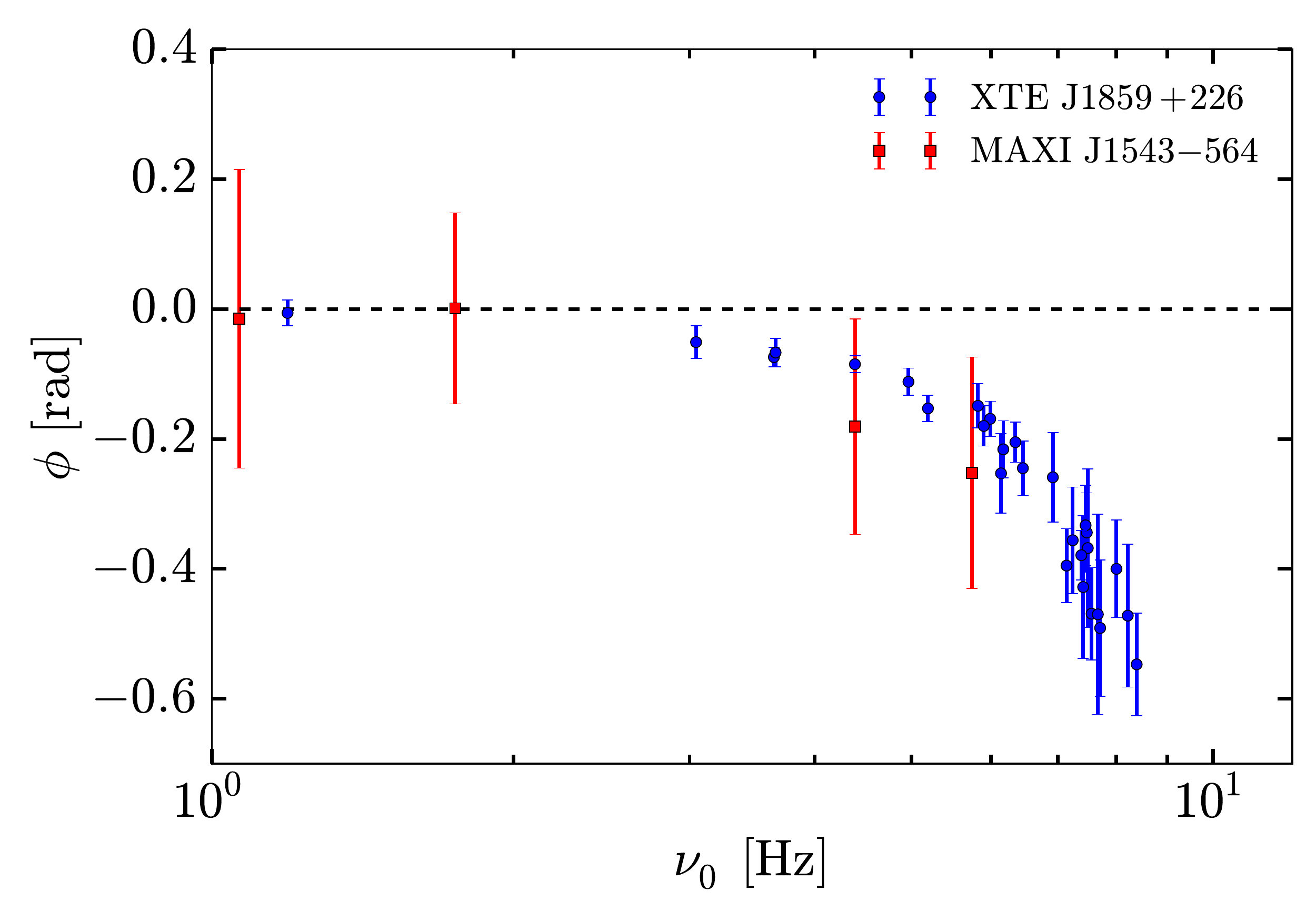}
    \caption{Type-C fundamental phase lags as a function of QPO frequency for XTE J1859+226 and MAXI J1543-564, both sources of undetermined inclination. Both sources show significant hard lags, consistent with the high inclination sample.}
    \label{fig:unknown}
  \end{center}
\end{figure}

For both XTE J1859+226 and MAXI J1543-564, no definite classification into low or high inclination could be determined. Thus, both sources where excluded from the comparison between the high and low inclination samples. In Figure \ref{fig:unknown}, we plot the Type-C QPO phase lag as a function of QPO frequency for these two sources. XTE J1859+226 clearly follows the same trend as the high inclination sources. Due to the small number of observations and larger uncertainties, the MAXI J1543-564 lags are less evident. However, the lags are clearly soft, consistent with the high inclination sample. Based on the QPO and noise amplitudes, \citet{motta15} also found that XTE J1859+226 and MAXI J1543-564 are most consistent with a high inclination. The clear presence of QPO frequency differences in XTE J1859+226 (Figure \ref{fig:df_inc}), previously only found in three high inclination sources, reinforces the interpretation that XTE J1859+226 is indeed a high inclination source.  

\subsubsection{GRO J1655-40}
\label{sec:gro}
\begin{figure}
  \begin{center}
    \includegraphics[width=\columnwidth]{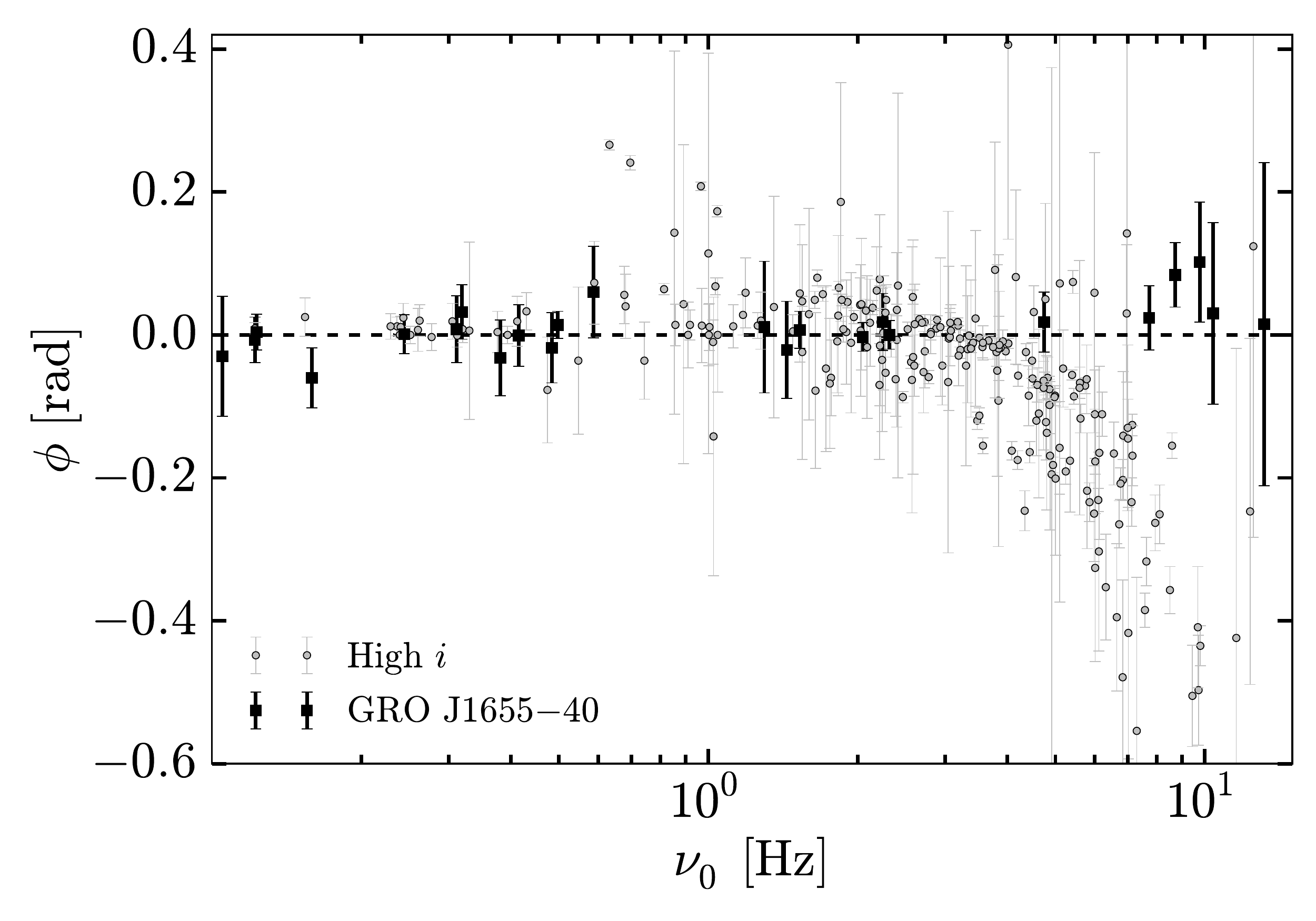}
    \caption{Type-C fundamental phase lags as a function of QPO frequency for GRO J1655-40 (black) and the remaining high inclination sources (grey). GRO J1655-40 shows constant phase lags, consistent with zero, contrary to the hard lags in other high inclination sources. This difference could be linked to the presence of a slim disc partially obscuring the precessing region (see text for details).}
    \label{fig:groj1655}
  \end{center}
\end{figure}

The high-inclination source GRO J1655-40 does not seem to show any significantly non-zero phase lags at the Type-C QPO frequency. To demonstrate this, we plot these phase lags versus QPO frequency, together with all other high inclination sources in grey, in Figure \ref{fig:groj1655}; GRO J1655-40's behaviour is clearly distinct from the other sources above approximately $1$ Hz. However, we emphasize that the reported inclination dependence remains significant despite the presence of GRO J1655-40 in our high inclination sample. Despite this note, and given the remarkable timing properties of this source in general, it is interesting to discuss the possible origin of its distinct lag behaviour.

Recently, \citet{uttley15} reported the detection of a \textit{hypersoft} state in GRO J1655-40, which is interpreted as the sign of a slim disk. As jet measurements indicate an extremely high inclination for its inner disk ($85 ^{\rm o}$, \citealt{hjellming95}), the inner accretion flow is possibly obscured by the slim disk. This obscuration could, in the Lense-Thirring precession model, lead to the QPO variations being dominated by quasi-periodic variations of the covering fraction of the inner flow by the slim disk, leading to simple flux variations without corresponding spectral changes that produce the lags. Such a scenario would also account for the other lags measured for GRO J1655-40, which are consisent with zero as well. In addition, it can account for the differences between GRO J1655-40 and GRS 1915+105: while both show remarkable timing properties, and possibly accrete at high rates \citep{neilsen16}, the inclination of GRS 1915+105 is less extreme ($70 ^{\rm o}$, \citealt{mirabel94}). Thus, in GRS 1915+105, nonzero lags are observed as the inclination is not high enough to obscure the inner flow.

\subsubsection{GRS 1915+105}
\label{sec:grs}

GRS 1915+105 was not considered by \citet{motta15} and \citet{heil15}, as it does not follow a standard q-shaped hardness-intensity diagram and possesses unique timing properties \citep{belloni10}. We have included this source in our sample, both because of the peculiar log-linear dependence of QPO phase lag on QPO frequency, and the clear presence of QPO frequency differences. Removing GRS 1915+105 decreases the signifance of the inclination dependence of the Type-C QPO phase lag: the p-value with the $1$ Hz cut increases to $p=0.0138$. This is both the result of the small uncertainties on the QPO phase lags in GRS 1915+105, and most importantly the small number of sources in our sample. Hence, removing a source with relatively many, well-constrained lag measurements has a clear impact on the significance of the inclination dependence. In fact, the removal of any source with at least as many Type-C QPO observations as GRS 1915+105 results in similar decreases in significance. In other words, the higher p-value appears to follow from the smaller sample size.

We find that, while the frequency differences are present in a selection of other sources, the log-linear phase lag behaviour is not seen in any other source. This is especially interesting towards the lowest QPO frequencies, where all other sources tend to zero phase lags, while the hard phase lags in GRS 1915+105 continue increasing. In our interpretation of the inclination dependence, this corresponds to a process where the flux increases as the precessing flow increases its inclination. For low inclination sources, at high QPO frequency, we suggested that this process could be caused by Doppler effects. However, this is less expected at the lower QPO frequencies where the QPO lags in GRS 1915+105 are hard, as this corresponds to a larger truncation radius. This is also shown by the model of \citet{veledina13}: our sketch of their results (see Figure \ref{fig:valedina}) shows that the influence of these Doppler effects becomes less prominent for precessing rings of larger radii. 

An alternative interpretation is proposed in \citet{vandeneijnden16} through differential precession. GRS 1915+105 is the only known source showing negative frequency differences (for QPO frequencies below $2$ Hz) \citep{qu10,yan12}. In the differential precession model, this is interpreted as a QPO frequency that increases towards the inner radius, spectrally weighted by emission that becomes softer towards smaller radii. In this model, hard lags could correspond to the average lag of the slower precessing, harder regions with respect to the faster precessing, softer regions. This interpretation could explain why only GRS 1915+105 shows hard lags at low QPO frequencies, as it is the only source showing negative frequency differences. However, why the hardness would decrease towards smaller radii for low QPO frequencies only in GRS 1915+105, remains unanswered. 

The log-linear dependence of QPO phase lag on frequency, including a sign change, was first reported by \citet{reig00}. There, the change in lag sign was interpreted as the result of a change in optical depth of the Comptonizing region: a low optical depth, assumed to be present at lower count rates and thus low QPO frequency, results in a hard lag. Higher optical depths, present at high count rates i.e. high QPO frequency, lead to a soft lag. This scenario does not assume a specific model for the QPO, and hence would be consistent with geometric QPO models. However, the difference with other high inclination sources, not showing hard lags, is not directly accounted for. 


\subsection{QPO frequency differences and differential precession}

We observe a significant non-zero frequency difference $\Delta \nu_0$ in four out of fifteen sources in our sample. In all cases, the trend as a function of QPO frequency is comparable, with a sudden increase in $\Delta \nu_0/\nu_0$ at the highest QPO frequencies ($> 6$--$7$ Hz). As already stated, three of these sources (GRS 1915+105, XTE J1550-564 and H1743-322) are high inclination. As discussed in Section \ref{sec:undetermined}, the fourth source, XTE J1859+226, shows very similar Type-C QPO lag behaviour to the high inclination sample. No low inclination source shows a non-zero $\Delta \nu_0$; XTE J1650-500 for instance covers the same range in QPO frequencies and phase lags (with opposite sign) as XTE J1859+226, while its $\Delta \nu_0$ does not show non-zero values\footnote{See Appendix A for plots of $\Delta \nu_0$ versus $\nu_0$ for each source.}. This is not due to the size of the uncertainties in XTE J1650-500 -- these are much smaller than many of the observed non-zero values of $\Delta \nu_0$ in other sources. 

In the high inclination source sample, three sources do not show significant frequency differences: GRO J1655-40, 4U 1630-47 and MAXI J1659-152. As discussed in section \ref{sec:gro}, the timing properties of GRO J1655-40 might be significantly influenced by the presence of a slim disk. In the latter two sources, the lack of frequency differences could simply result from the data quality: for 4U 1630-47, only few QPO observations, with large phase lag uncertainties, are analysed. In MAXI J1659-152, the QPO frequency barely passes $7$ Hz, after which most non-zero frequency differences in other sources are observed. Hence, despite the lack of a clear inclination dependence, the frequency difference might only be observable in high inclination sources. The analysis of new transient BHXRBs or outbursts is thus vital to investigate this further.

Analysing $\Delta \nu_0$ sheds light on the short-time-scale evolution of phase lags in Type-C QPOs. Such lag evolution was reported by \citet{vandeneijnden16} in GRS 1915+105 and interpreted as a sign of differential precession of the inner accretion flow. In this interpretation, different radii in the inner flow precess at different rates, resulting in the gradual build-up of a warp over the course of $5-10$ precession cycles. This warp in turn destabalises the precession, leading to the decoherence of the QPO. This process results in the presence of an enveloping structure (\textit{coherent intervals)} in the QPO amplitude, in which the phase lag systematically increases, and might be related to the Q-factor and \textit{quasi}-periodic nature of QPOs. Although not all sources show a significant $\Delta \nu_0$, the coherent intervals appear to be present in all sources, as expected in this interpretation: even if no $\Delta \nu_0$ is observed, the precession is still destabilised by the its differential nature. 

We measure $\Delta \nu_0$ as a probe of this effect, as the actual measurement of evolving phase lag requires too high a signal-to-noise to be applicable to all sources in our sample. This approach thus assumes that the link between differential precession and phase lag evolution is generally applicable to BHXRBs. A more problematic downside of this method is that the observed $\Delta \nu_0$ is not a direct proxy of the intrinsic difference in precession rate between different radii in the inner accretion flow: we can only observe the spectrally weighted $\Delta \nu_0$. If each radius emits with the same spectral shape, all energy bands will consist of the same weighted combination of precession frequencies. Thus, no $\Delta \nu_0$ will be observed. 

As any observed $\Delta \nu_0$ is always spectrally weighted, investigating its inclination dependence could provide further information on the emission processes of the differentially precessing flow. If frequency differences are indeed only visible in edge-on sources, this would imply that the inner flow's spectral appearance in edge-on sources generally differs more at different radii than in face-on sources. This could for instance result from differences in the amount of scatterings between different inclinations: in face-on sources, photons received from all radii have scattered approximately the same amount of times. In edge-on sources, the innermost regions might be more obscured. As a result, photons from the smallest radii have scattered more, resulting in a harder observed spectrum from these smaller radii. To test such a scenario, it is important to test the hypothesis that Type-C QPO frequency differences are only visible at high inclination, in a new sample of sources.

\section{Conclusions}
\label{sec:conclusion2}
In this paper, we present the results of a systematic investigation of the inclination dependence of both phase lags and QPO frequency differences in BHXRBs. We conclude that
\begin{enumerate}
    \item the Type-C QPO lag depends clearly on source inclination, both in sign and behaviour as a function of QPO frequency,
    \item The Type-B QPO lag does not depend significantly on source inclination, but shows distinct behaviour as a function of QPO frequency compared to the Type-C QPO,
    \item the Type-C BBN lags show a dependence on inclination, but no clear dependence on QPO frequency,
    \item four BHXRBs show a difference in Type-C QPO frequency between the hard and soft band, none of which are of low inclination,
    \item XTE J1859+226, and to a lesser extent MAXI J1543-564, are consistent with being high inclination sources.
\end{enumerate}
These results provide a clear indication that both the Type-B and Type-C QPOs originate from two distinct geometric processes, consistent with recent results. It is important that the inclination-dependence hypothesis, both for the lags and QPO frequency differences, will be tested in the future in new sources and/or outbursts. The Type-C QPO lags show a very clear and simple difference between high and low inclination: the sign of the phase lag. Hence, this will be easy to test with observations of new black hole transients which show Type-C QPOs above a few Hz and for which the inclination can be determined. Furthermore, systematically investigating energy-dependent frequency differences in the Type-C harmonic or Type-B fundamental could provide new insights in the differences between QPO types and harmonics. 

\section*{Acknowledgements}
We thank the anonymous referee for insightful comments that improved the quality of this work. This research has made use of data obtained through the High Energy Astrophysics Science Archive Research Center Online Service, provided by the NASA/Goddard Space Flight Center. AI acknowledges support from the Netherlands Organization for Scientific Research (NWO) Veni Fellowship, grant number 639.041.437. TMB acknowledges support from grant ASI-INAF I/037/12/0. SEM acknowledges the University of Oxford and the Violette and Samuel Glasstone Research Fellowship program.




\bibliographystyle{mnras}
 \newcommand{\noop}[1]{}



\appendix

\section{Additional plots for individual sources}

\begin{figure}
  \begin{center}
    \includegraphics[width=\columnwidth]{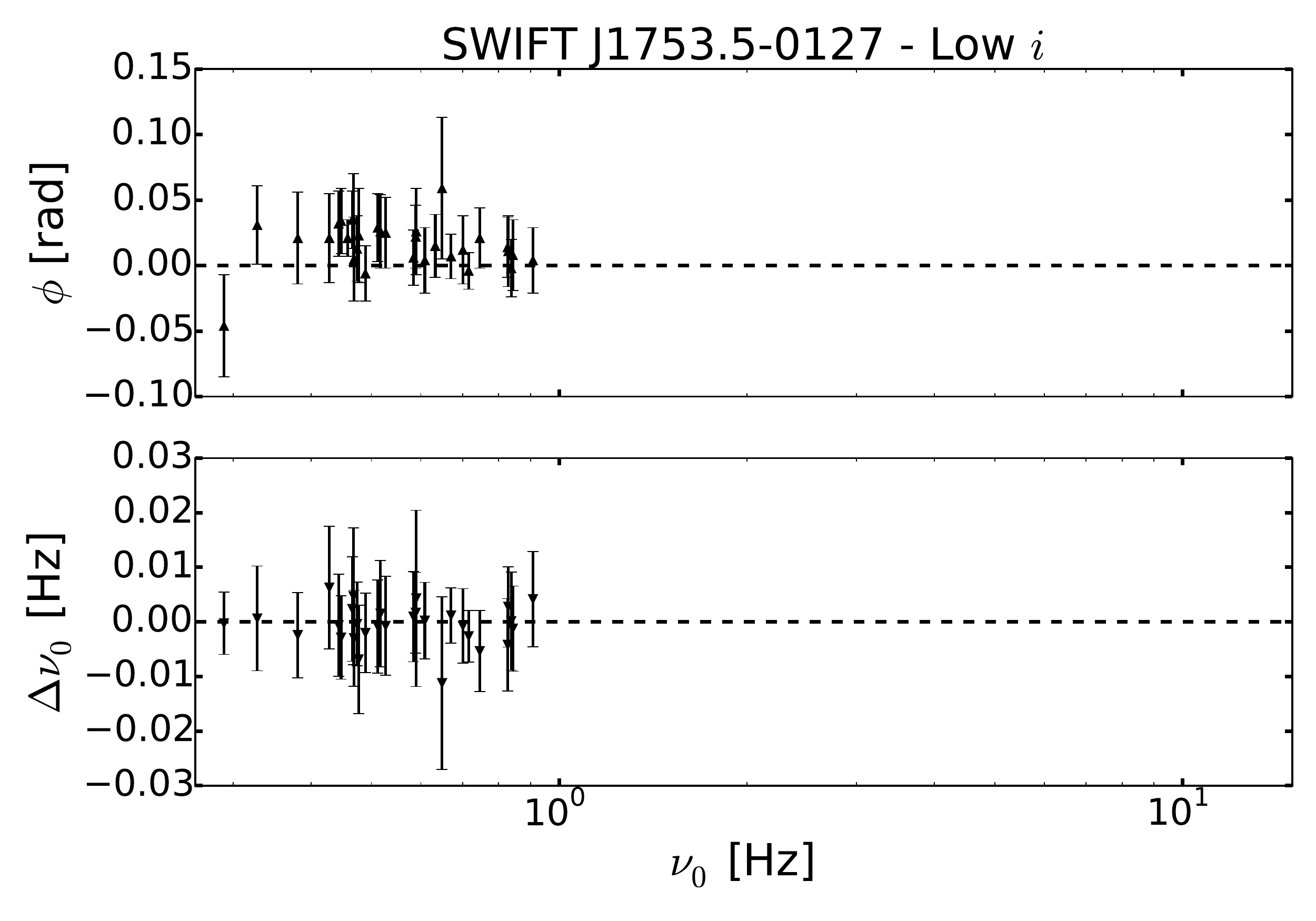}
    \caption{Type-C QPO phase lag and $\Delta \nu_0/\nu_0$ as a function of QPO frequency for SWIFT J1753.5-0127}
    \label{fig:df_and_phase_SWIFT_J1753.5-0127}
  \end{center}
\end{figure}

\begin{figure}
  \begin{center}
    \includegraphics[width=\columnwidth]{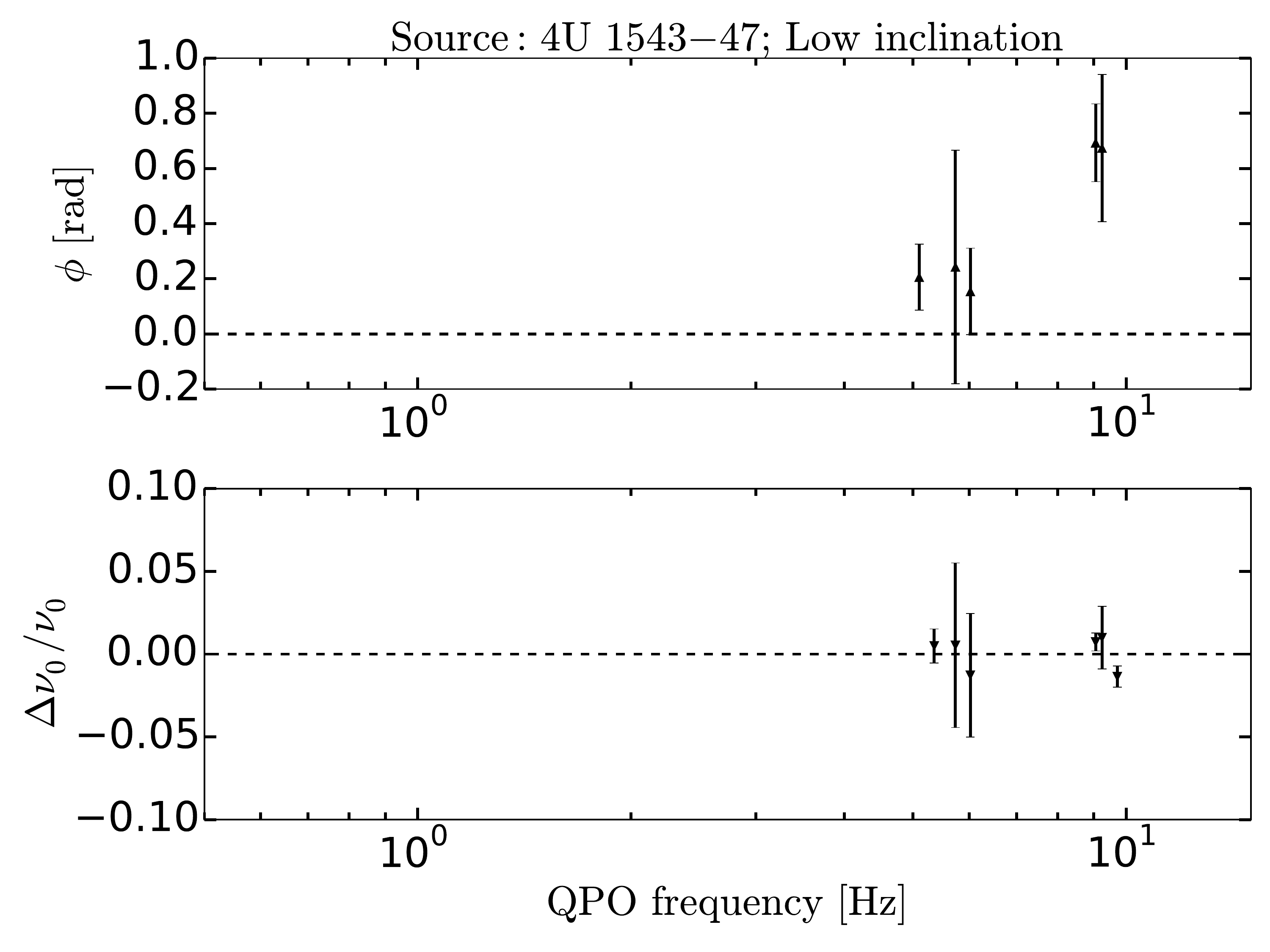}
    \caption{Type-C QPO phase lag and $\Delta \nu_0/\nu_0$ as a function of QPO frequency for 4U 1543-47}
    \label{fig:df_and_phase_4U1543-47}
  \end{center}
\end{figure}

\begin{figure}
  \begin{center}
    \includegraphics[width=\columnwidth]{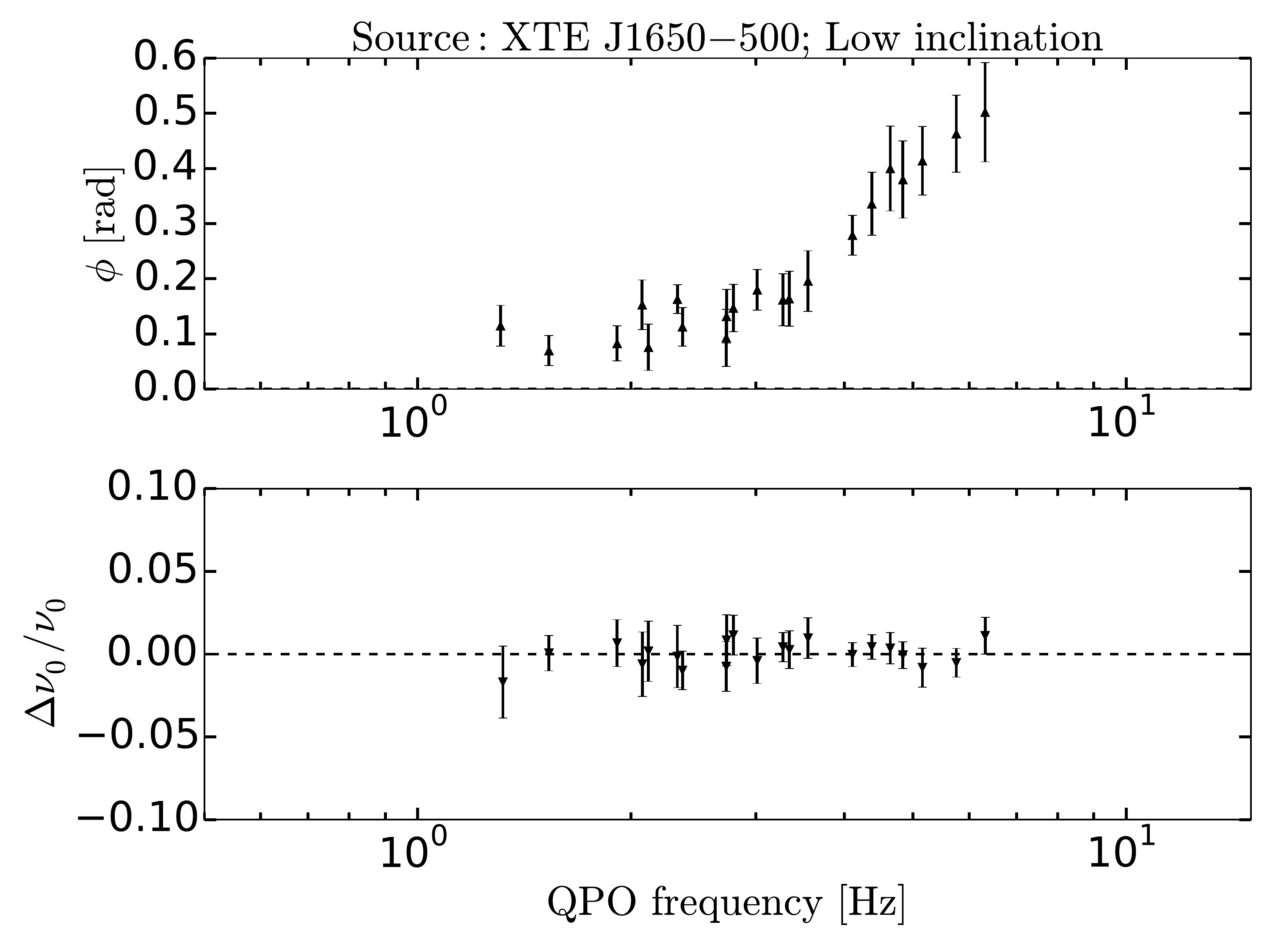}
    \caption{Type-C QPO phase lag and $\Delta \nu_0/\nu_0$ as a function of QPO frequency for XTE J1650-500}
    \label{fig:df_and_phase_XTE_J1650-500}
  \end{center}
\end{figure}

\begin{figure}
  \begin{center}
    \includegraphics[width=\columnwidth]{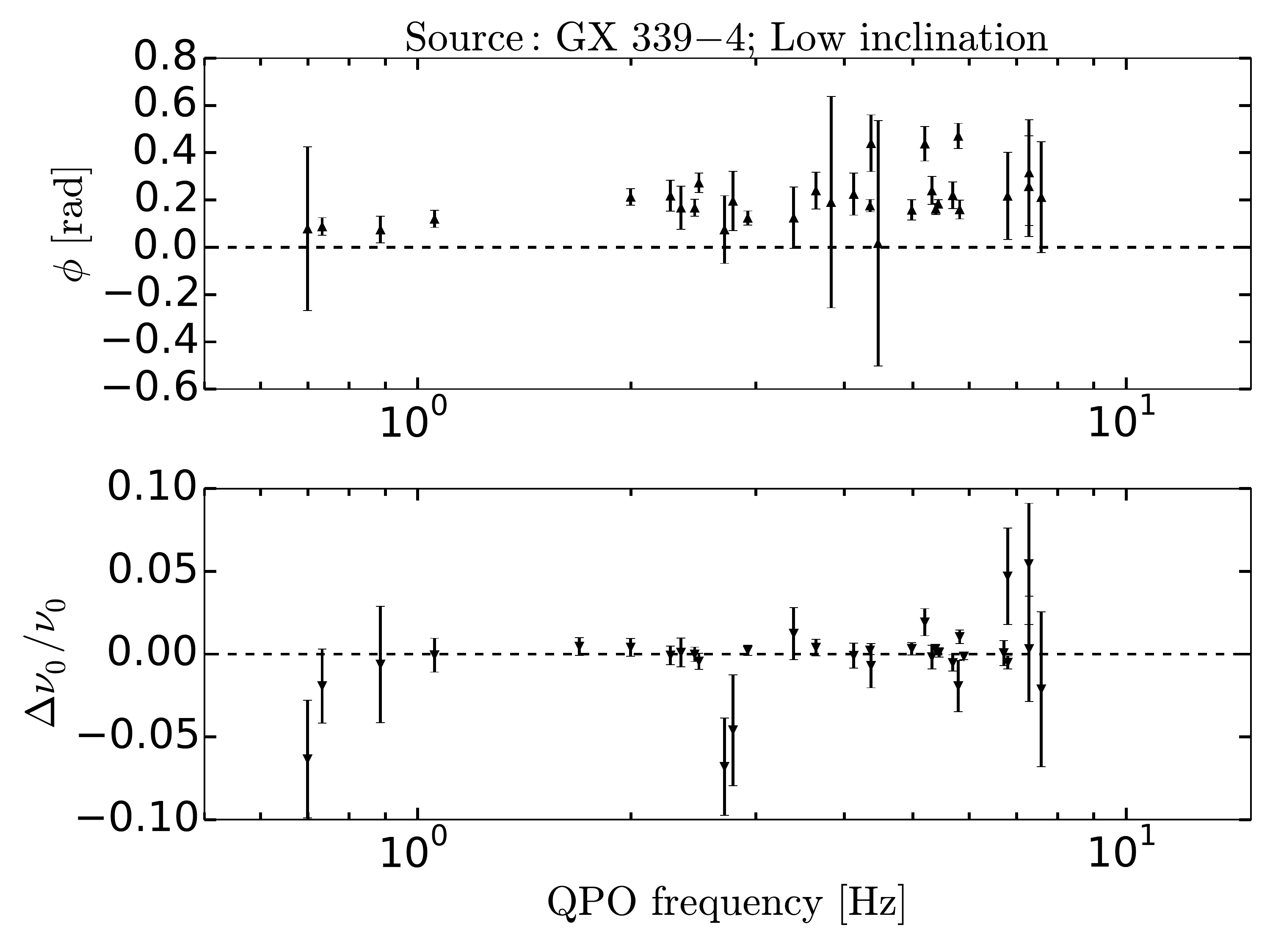}
    \caption{Type-C QPO phase lag and $\Delta \nu_0/\nu_0$ as a function of QPO frequency for GX 339-4}
    \label{fig:df_and_phase_GX_339-4}
  \end{center}
\end{figure}

\begin{figure}
  \begin{center}
    \includegraphics[width=\columnwidth]{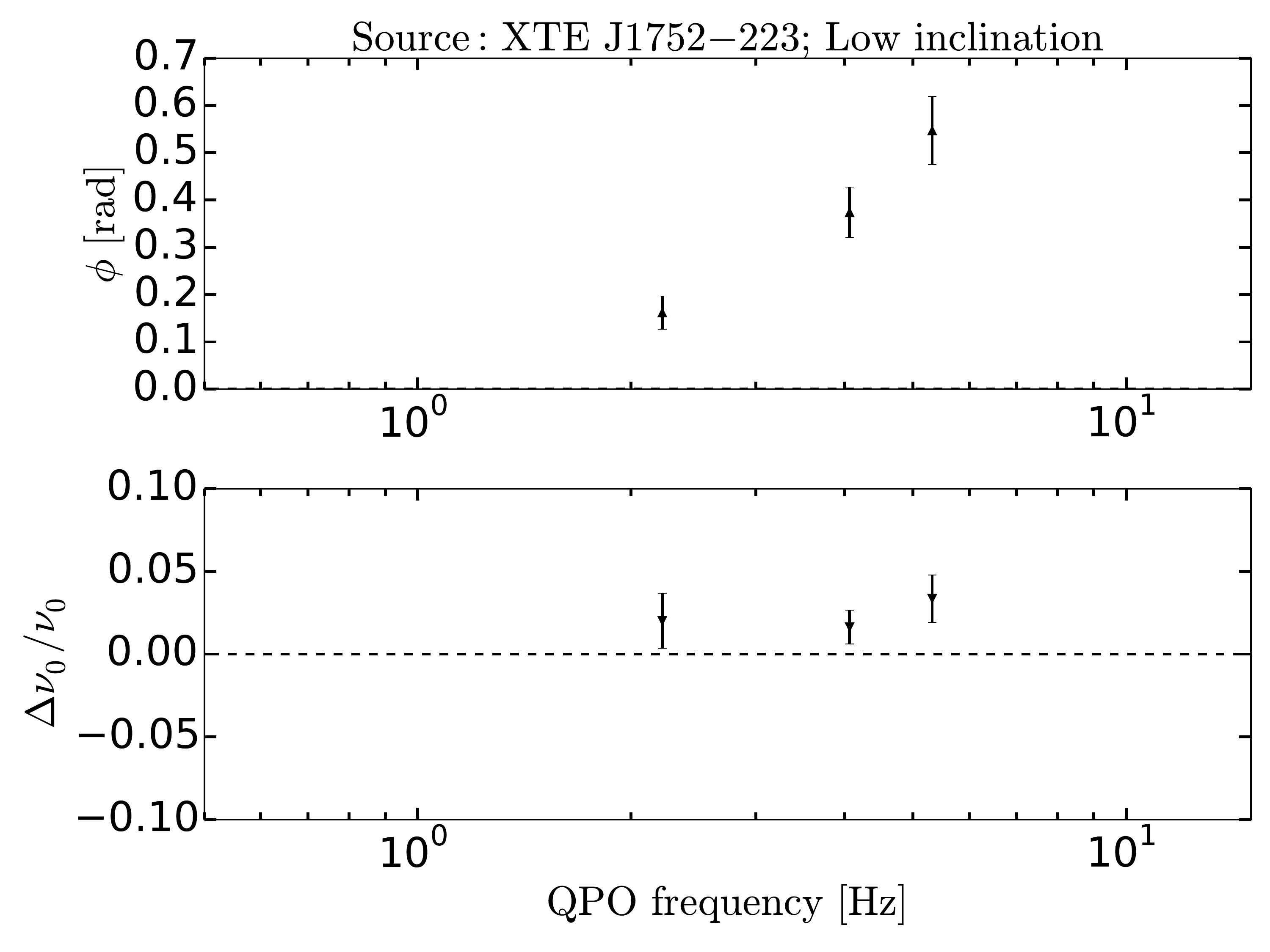}
    \caption{Type-C QPO phase lag and $\Delta \nu_0/\nu_0$ as a function of QPO frequency for XTE J1752-223}
    \label{fig:df_and_phase_XTE_J1752-223}
  \end{center}
\end{figure}

\begin{figure}
  \begin{center}
    \includegraphics[width=\columnwidth]{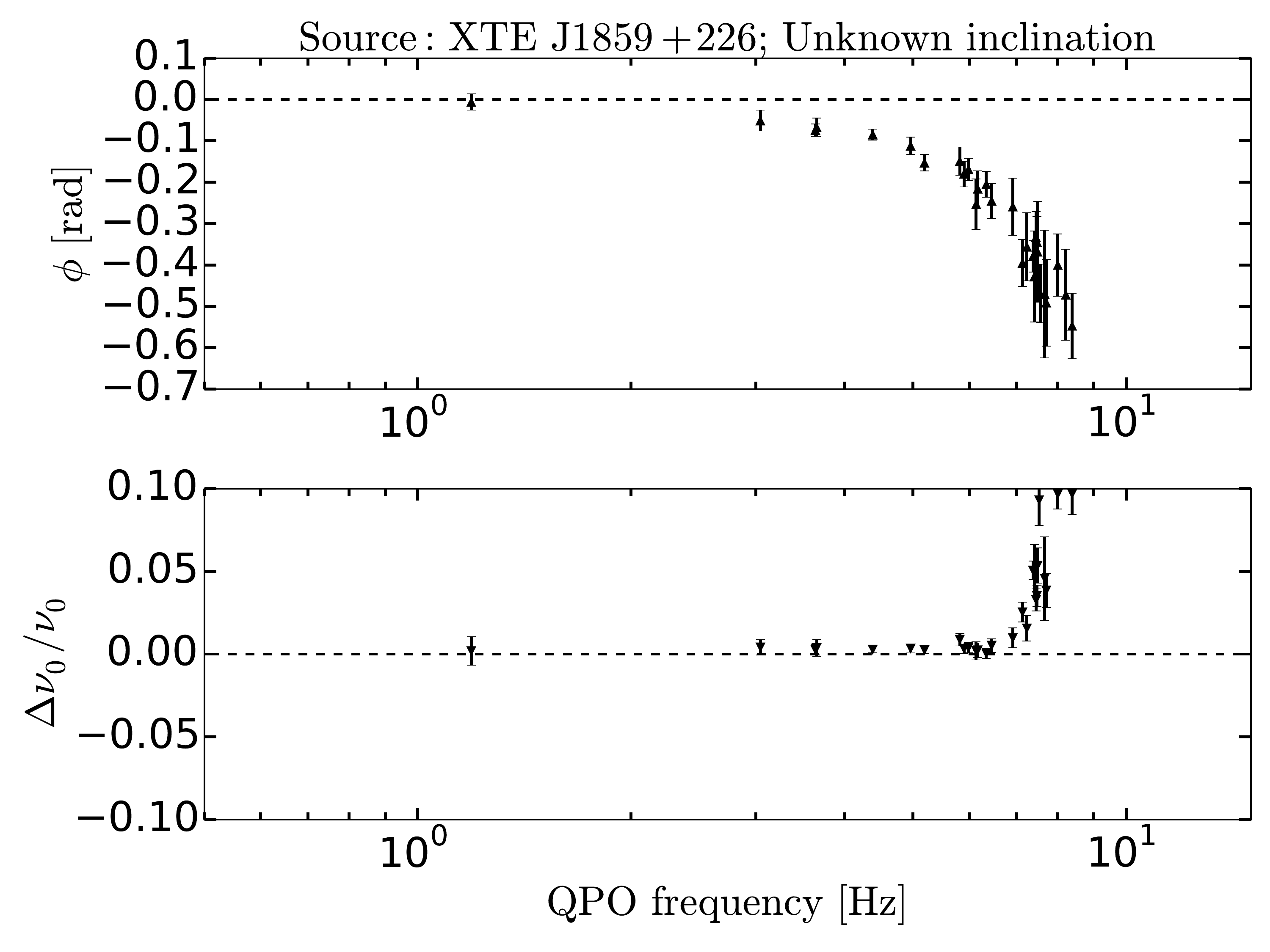}
    \caption{Type-C QPO phase lag and $\Delta \nu_0/\nu_0$ as a function of QPO frequency for XTE J1859+226.}
    \label{fig:df_and_phase_XTE_J1859+226}
  \end{center}
\end{figure}

\begin{figure}
  \begin{center}
    \includegraphics[width=\columnwidth]{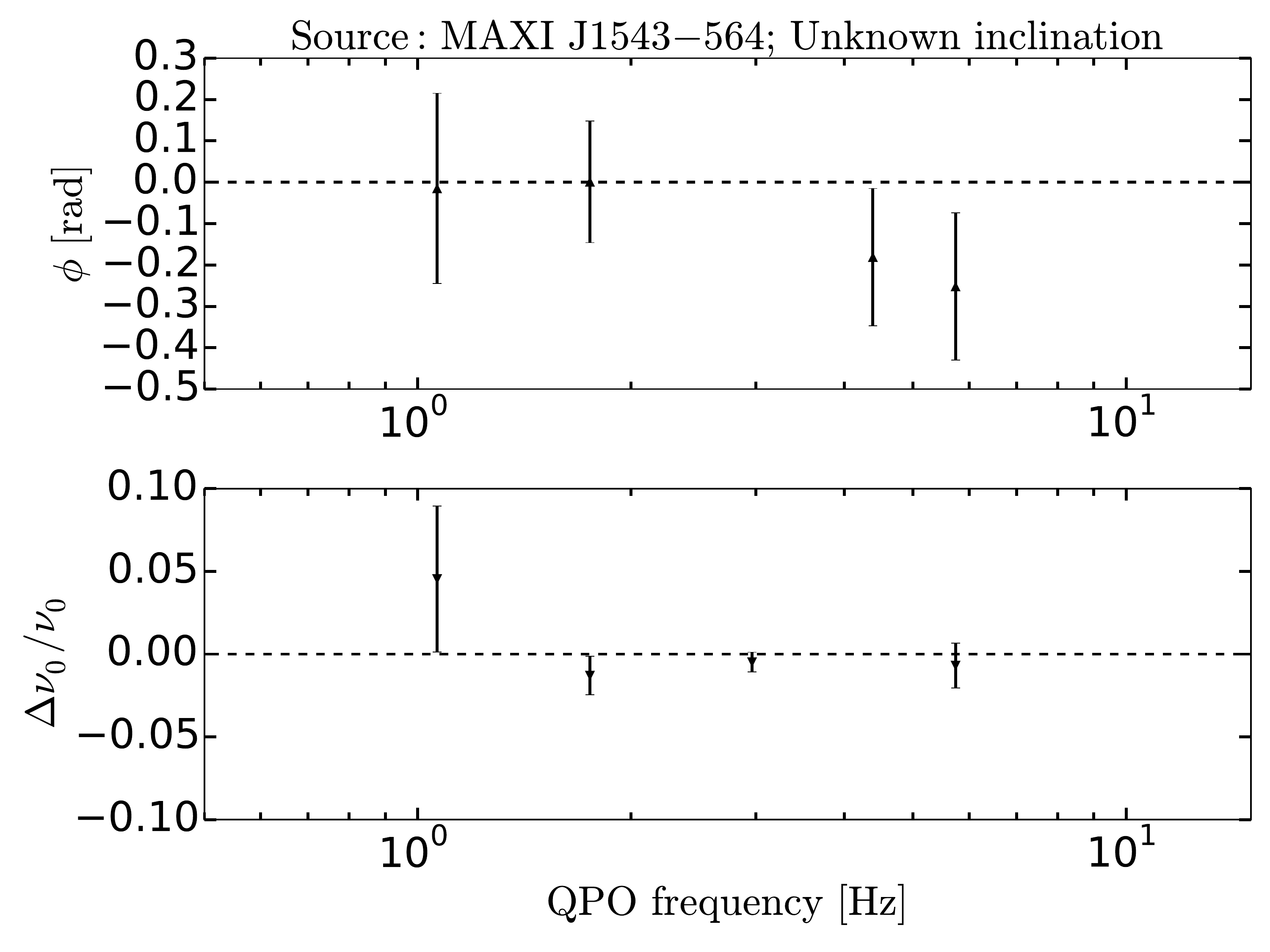}
    \caption{Type-C QPO phase lag and $\Delta \nu_0/\nu_0$ as a function of QPO frequency for MAXI J1543-564}
    \label{fig:df_and_phase_MAXI_J1543-564}
  \end{center}
\end{figure}

\begin{figure}
  \begin{center}
    \includegraphics[width=\columnwidth]{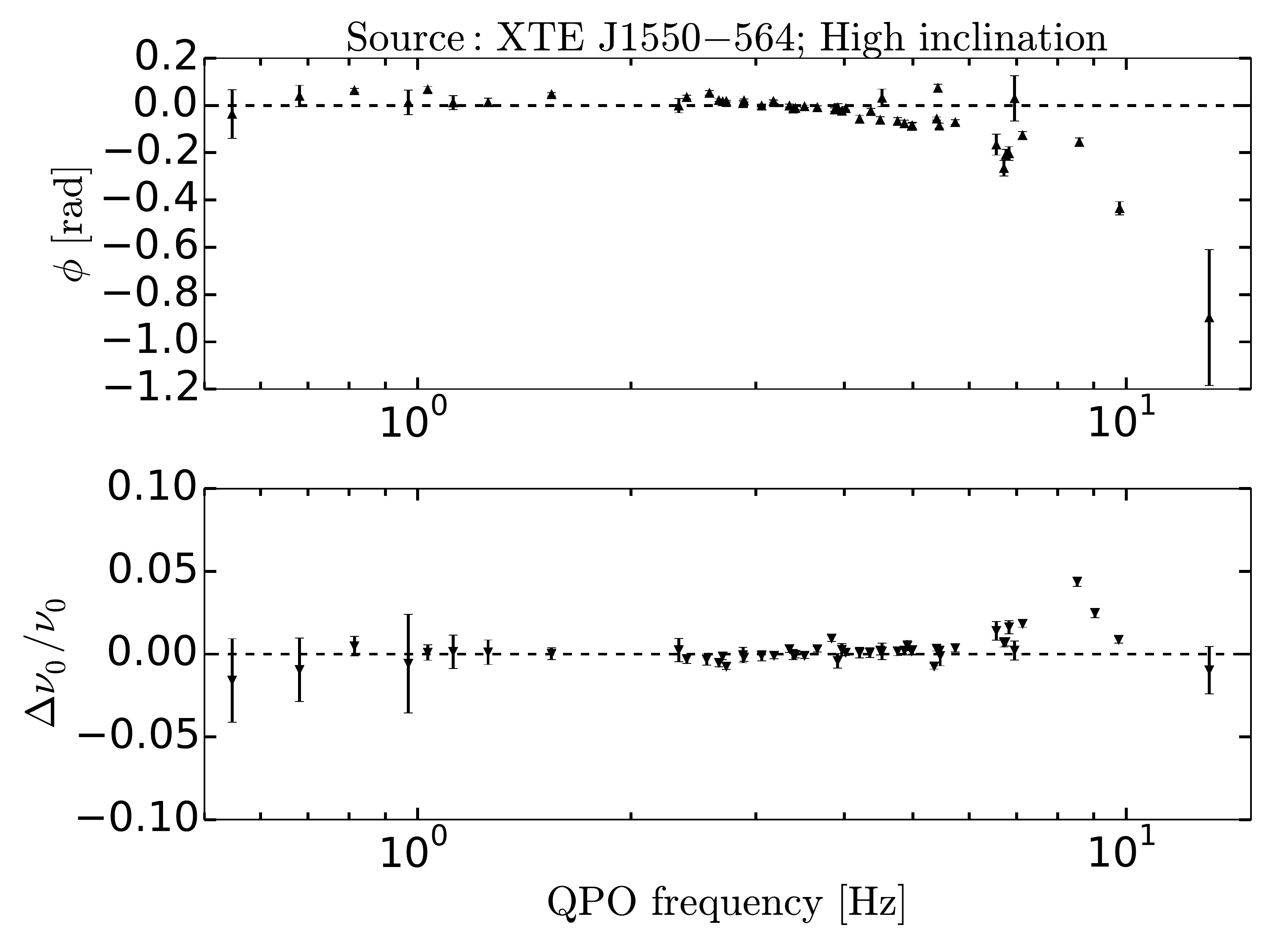}
    \caption{Type-C QPO phase lag and $\Delta \nu_0/\nu_0$ as a function of QPO frequency for XTE J1550-564.}
    \label{fig:df_and_phase_XTE_J1550-564}
  \end{center}
\end{figure}

\begin{figure}
  \begin{center}
    \includegraphics[width=\columnwidth]{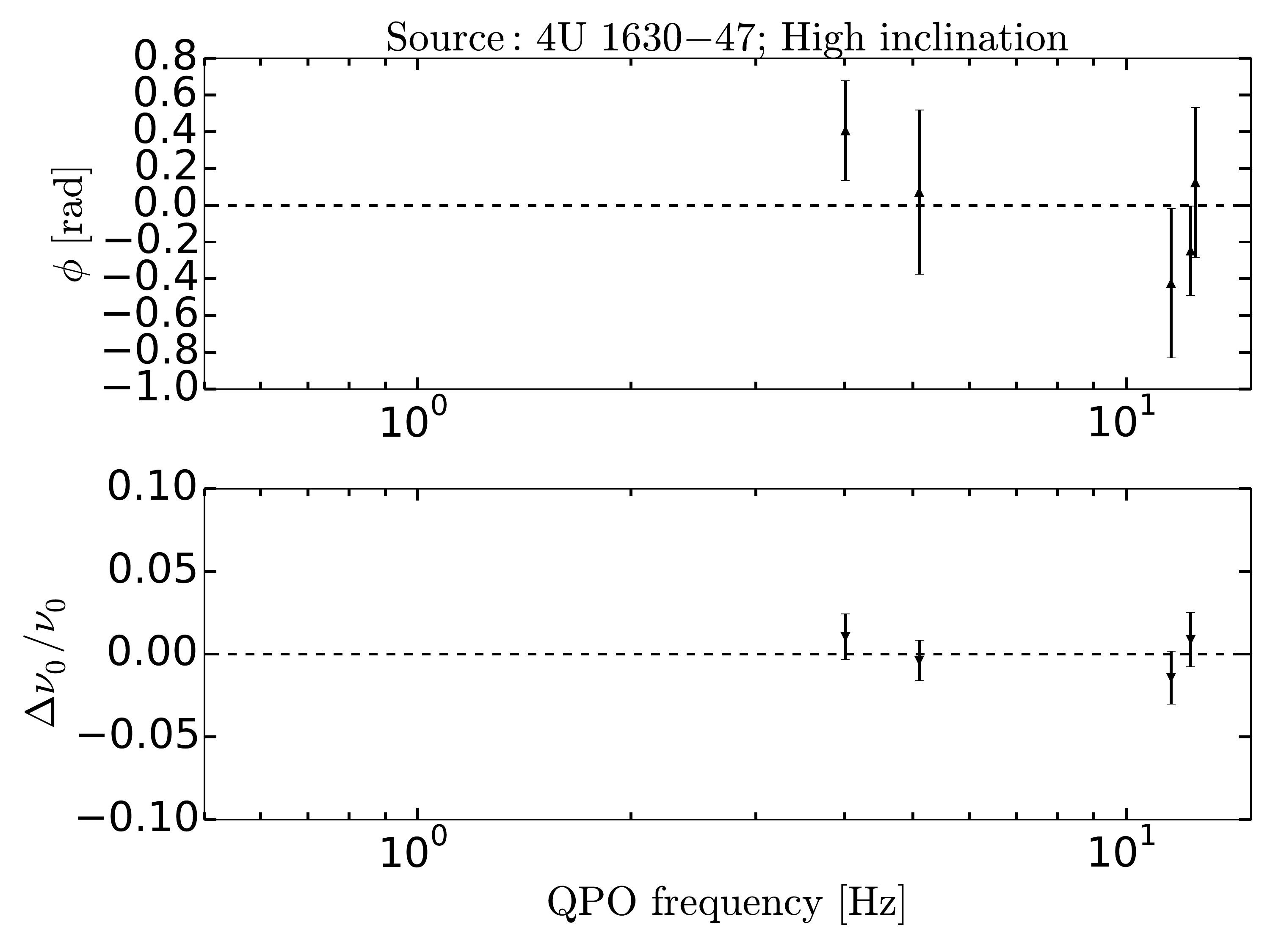}
    \caption{Type-C QPO phase lag and $\Delta \nu_0/\nu_0$ as a function of QPO frequency for 4U 1630-47}
    \label{fig:df_and_phase_4U_1630-47}
  \end{center}
\end{figure}

\begin{figure}
  \begin{center}
    \includegraphics[width=\columnwidth]{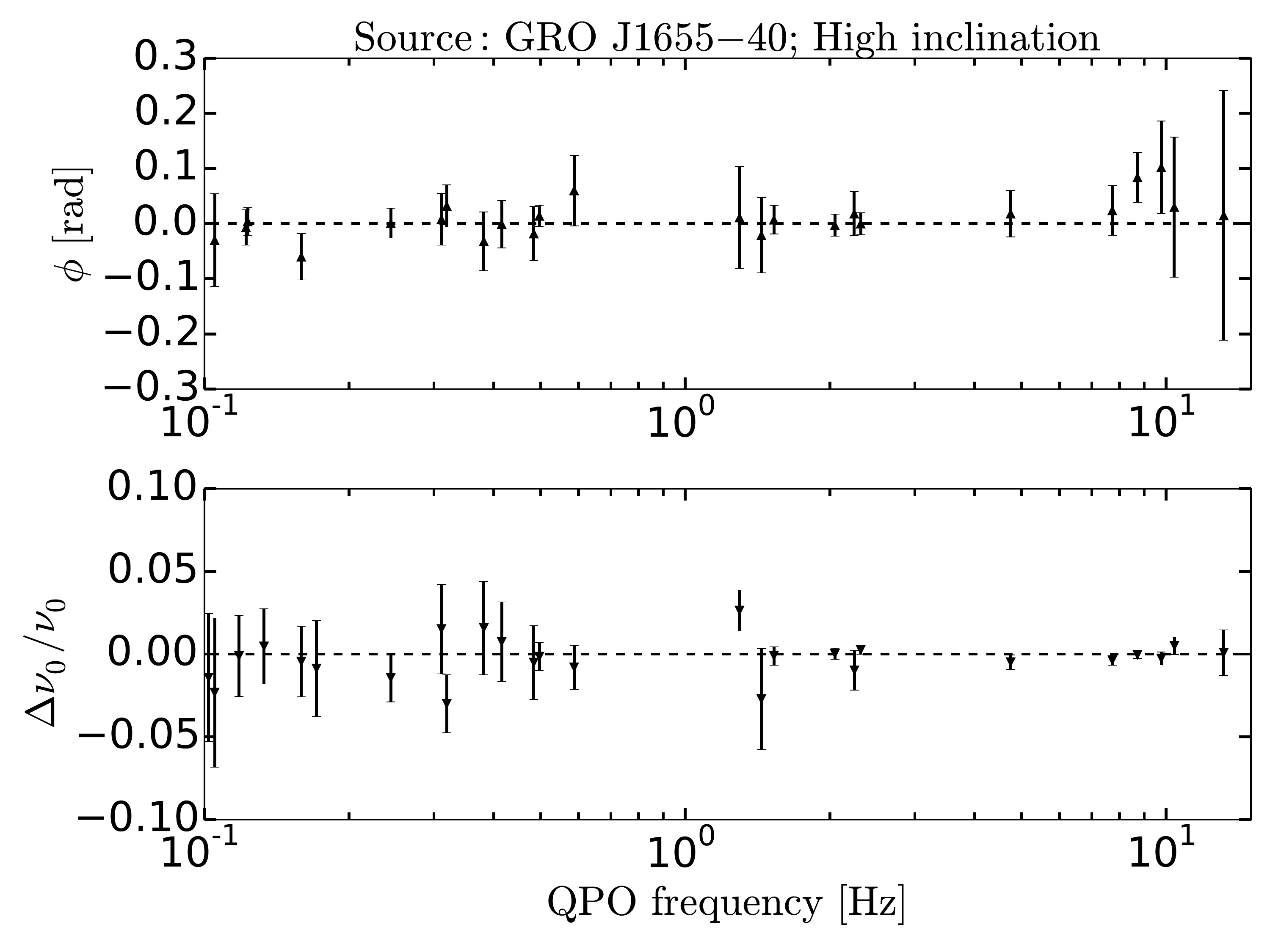}
    \caption{Type-C QPO phase lag and $\Delta \nu_0/\nu_0$ as a function of QPO frequency for GRO J1655-40}
    \label{fig:df_and_phase_GRO_J1655-40}
  \end{center}
\end{figure}

\begin{figure}
  \begin{center}
    \includegraphics[width=\columnwidth]{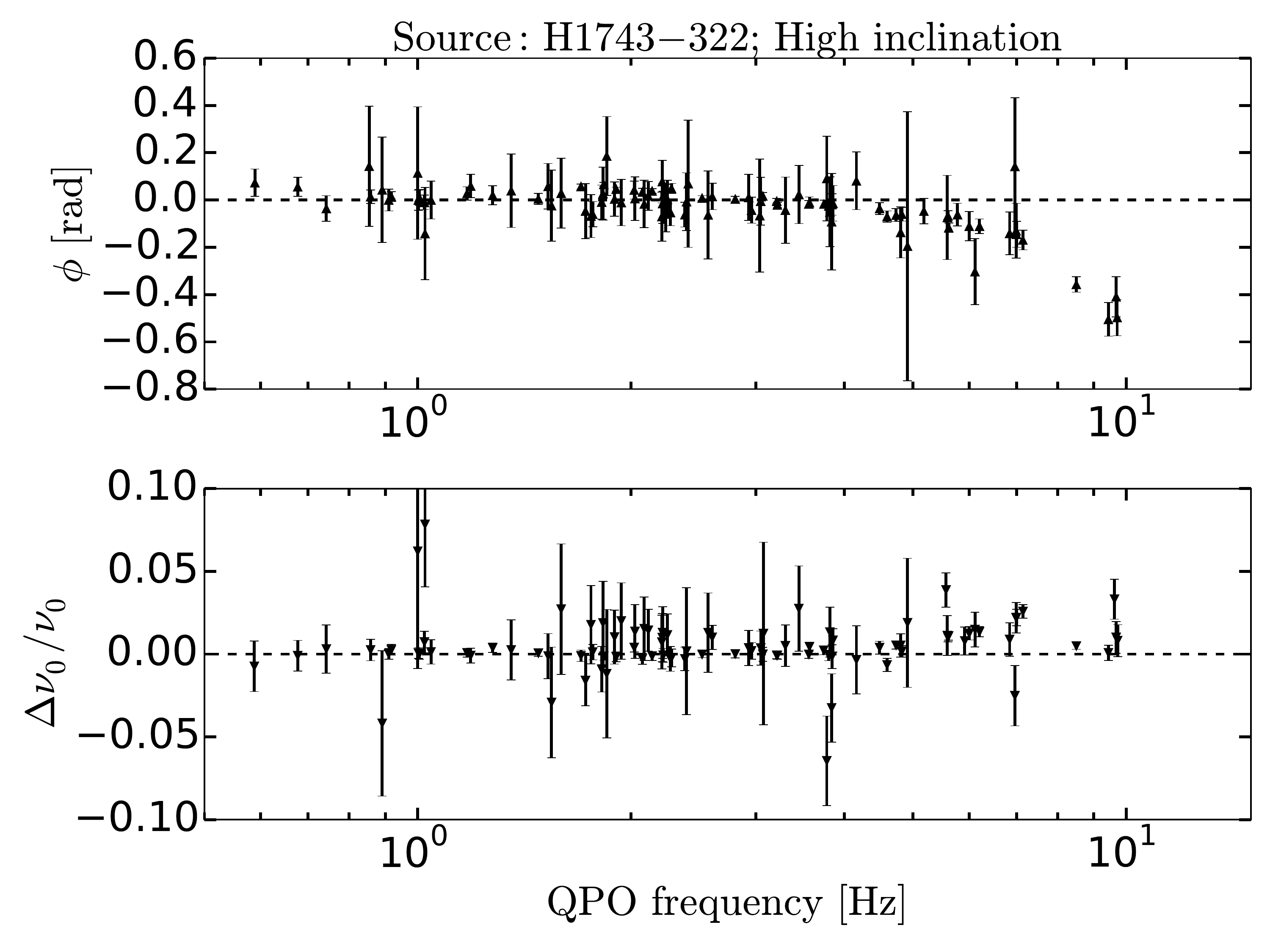}
    \caption{Type-C QPO phase lag and $\Delta \nu_0/\nu_0$ as a function of QPO frequency for H1743-322.}
    \label{fig:df_and_phase_XTE_J1746-319}
  \end{center}
\end{figure}

\begin{figure}
  \begin{center}
    \includegraphics[width=\columnwidth]{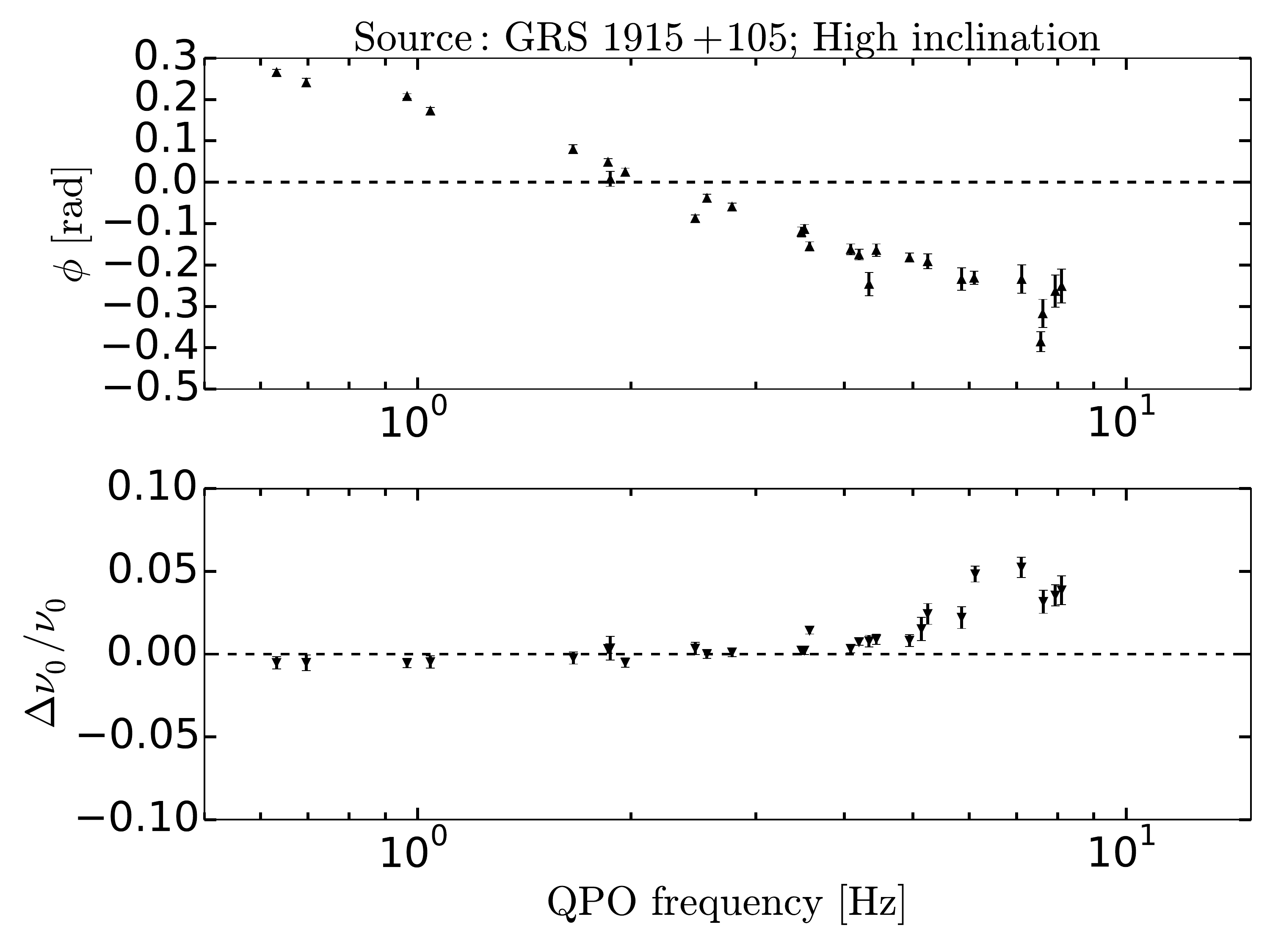}
    \caption{Type-C QPO phase lag and $\Delta \nu_0/\nu_0$ as a function of QPO frequency for GRS 1915+105.}
    \label{fig:df_and_phase_GRS1915}
  \end{center}
\end{figure}

\begin{figure}
  \begin{center}
    \includegraphics[width=\columnwidth]{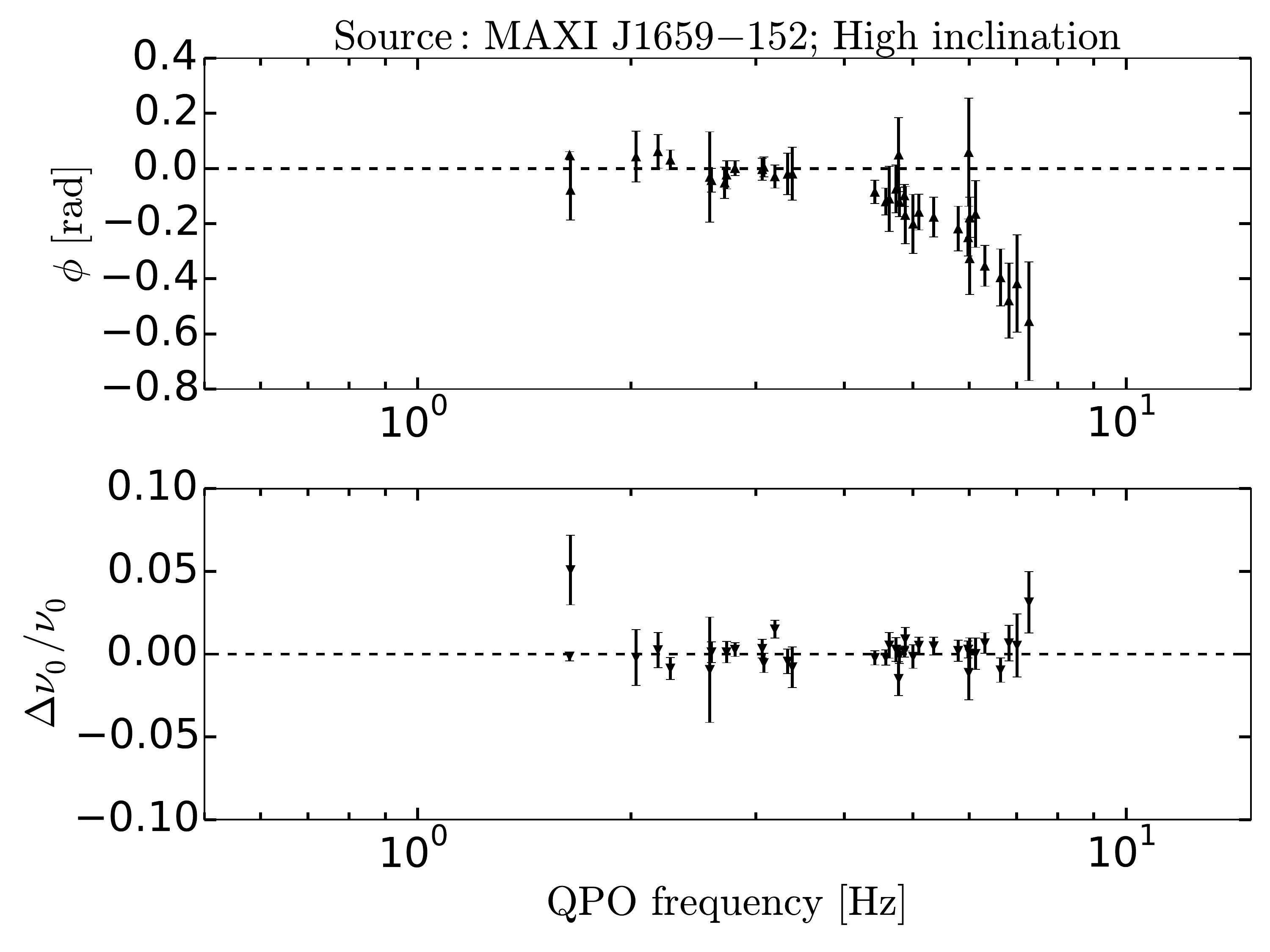}
    \caption{Type-C QPO phase lag and $\Delta \nu_0/\nu_0$ as a function of QPO frequency for MAXI J1659-152}
    \label{fig:df_and_phase_MAXI_J1659-152}
  \end{center}
\end{figure}

\begin{figure}
  \begin{center}
    \includegraphics[width=\columnwidth]{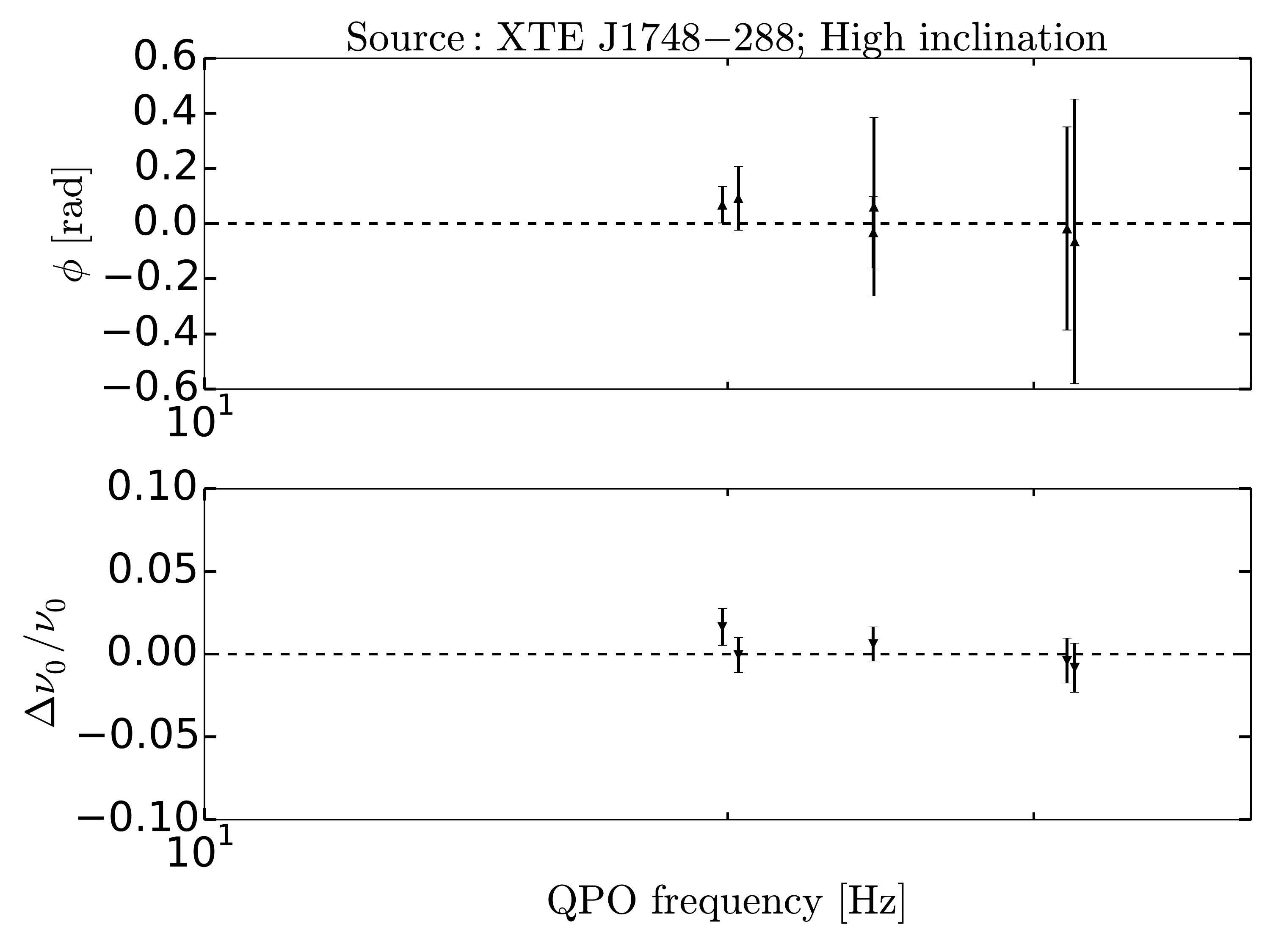}
    \caption{Type-C QPO phase lag and $\Delta \nu_0/\nu_0$ as function of QPO frequency for XTE J1748-288}
    \label{fig:df_and_phase_XTE_J1748_288}
  \end{center}
\end{figure}

In Figure \ref{fig:df_and_phase_SWIFT_J1753.5-0127} -- \ref{fig:df_and_phase_XTE_J1748_288}, we plot the Type-C fundamental phase lag and the relative QPO frequency difference as a function of QPO frequency for all individual sources showing Type-C QPOs. This excludes the source XTE J1817-330, for which only Type-B QPOs were analysed.

\section{Table of observations}

The online material contains an overview of analysed Type-C and Type-B QPO observations, respectively. Example rows of these Tables are shown in Table \ref{tab:example}. The online tables list the RXTE ObsID, the QPO frequency $\nu_{\rm QPO}$, the phase lag at the QPO fundamental, harmonic, subharmonic and BBN ($\phi_{\rm QPO}$, $\phi_{\rm harm}$, $\phi_{\rm sub}$, and $\phi_{\rm BBN}$, respectively), and the QPO frequency difference $\Delta \nu_0$. (Sub)harmonic lags are only shown when a (sub)harmonic is fitted in the power spectrum. Missing QPO lags, BBN lags or frequency differences indicate that this the uncertainty in the determination of this value was too large due to low signal-to-noise. 
\begin{table*}
 \begin{center}
  \caption{\small{Overview of analysed QPO observations. Full table available online.}}
  \label{tab:example}
   \begin{tabular}{lcccccc}
  ObsID & $\nu_{\rm QPO}$ [Hz] & $\phi_{\rm QPO}$ [rad] & $\phi_{\rm harm}$ [rad] & $\phi_{\rm sub}$ [rad] & $\phi_{\rm BBN}$ [rad] & $\Delta \nu_0$ [Hz] \\
  \hline \hline 
  \multicolumn{7}{c}{\textbf{XTE J1859+226 -- Undetermined inclination}} \\
  \hline \hline
40124-01-05-00 & $3.045$ & $-0.051 \pm 0.025$ & $0.127 \pm 0.068$ & $-0.029 \pm 0.051$ & $-0.029 \pm 0.055$ & $0.013 \pm 0.014$ \\ 
40124-01-06-00 & $3.642$ & $-0.074 \pm 0.015$ & $0.082 \pm 0.047$ & $-0.057 \pm 0.031$ & $-0.054 \pm 0.034$ & $0.009 \pm 0.012$ \\ 
40124-01-07-00 & $3.656$ & $-0.067 \pm 0.022$ & $0.108 \pm 0.068$ & $-0.069 \pm 0.044$ & $0.0 \pm 0.057$ & $0.014 \pm 0.019$ \\ 
40124-01-08-00 & $4.388$ & $-0.085 \pm 0.013$ & $0.116 \pm 0.044$ & $-0.054 \pm 0.025$ & $-0.064 \pm 0.027$ & $0.012 \pm 0.009$ \\ 
40124-01-09-00 & $4.963$ & $-0.112 \pm 0.021$ & $0.174 \pm 0.091$ & $-0.074 \pm 0.038$ & $-0.065 \pm 0.035$ & $0.017 \pm 0.011$ \\ 
\hline  
  \end{tabular}
  \end{center}
\end{table*}



\bsp	
\label{lastpage}
\end{document}